\documentclass[fleqn,usenatbib]{mnras}

\usepackage[T1]{fontenc}
\usepackage{ae,aecompl}

\usepackage{hyperref}
\usepackage{booktabs}
\usepackage{multirow}
\usepackage{pgf}
\usepackage{subcaption}
\usepackage{threeparttable}

\usepackage{graphicx}	% Including figure files
\usepackage{amsmath}	% Advanced maths commands
\usepackage{amssymb}	% Extra maths symbols

\usepackage{newtxtext,newtxmath}
\newcommand{\angs}{\textup{\AA}}
\newcommand{\DPS}{${\Delta\mbox{PS}}$}
\newcommand{\DMS}{${\Delta\mbox{MS}}$}

\newcommand{\lRe}{\lambda_{R_{\rm e}}}
\newcommand{\Ha}{\ion{H}{$\;\!\!\alpha$}}
\newcommand{\Hb}{\ion{H}{$\;\!\!\beta$}}
\newcommand{\Oiii}{[\ion{O}{III}]}
\newcommand{\Nii}{[\ion{N}{II}]}
\newcommand{\Reff}{$R_{\mathrm e}$}

\newcommand{\citeapos}[1]{\citeauthor{#1}'s (\citeyear{#1})}

\defcitealias{Tous+2020}{TSP20}
\title[Activity profiles of S0 galaxies]{The local Universe in the era of large surveys - III.\ Radial activity profiles of S0 galaxies}

\author[J.\ L.\ Tous et al.]{
J.\ L.\ Tous$^{1,2}$\thanks{E-mail: jtous@fqa.ub.edu},
J.\ M.\ Solanes$^{1,2}$, J.\ D.\ Perea$^{3}$ and H. Dom\'\i nguez-S\'anchez$^{4}$\\
$^{1}$Institut de Ci\`encies del Cosmos (ICCUB), Universitat de Barcelona, Mart\'i i Franqu\`es 1, E-08028 Barcelona, Spain\\
$^{2}$Departament de F\'isica Qu\`antica i Astrof\'isica, Universitat de Barcelona, Mart\'i i Franqu\`es 1, E-08028 Barcelona, Spain\\
$^{3}$Departamento de Astronom\'ia Extragal\'actica, Instituto de Astrof\'isica de Andaluc\'ia, IAA-CSIC, Glorieta de la Astronom\'ia s/n, E-18008 Granada, Spain\\
$^{4}$Centro de Estudios de F\'\i sica del Cosmos de Arag\'on (CEFCA). Plaza San Juan, 1, E-44001, Teruel, Spain
}

\date{Accepted XXX. Received YYY; in original form ZZZ}

\pubyear{2023}

\begin{document}
\label{firstpage}
\pagerange{\pageref{firstpage}--\pageref{lastpage}}
\maketitle

\begin{abstract}
Spatially resolved MaNGA's optical spectra of 1072 present-day lenticular (S0) galaxies, dimensionally reduced from a principal component analysis (PCA), are used to determine their radial activity structure shaped by any possible nebular ionization source. Activity profiles within $1.5\,$\Reff\ are examined in tandem with the mass, age, ellipticity and kinematics of the stars, as well as environmental density. Among the results of this comparison, we find that the sign of the radial activity gradient of S0s is tightly related to their PCA classification, BPT designation, and star formation status. PCA-passive lenticulars often show low-level, flat activity profiles, although there is also a significant number of systems with positive gradients, while their less common active counterparts generally have negative gradients, usually associated with high SSFRs and, sometimes, moderate Seyfert emission. A fraction of the latter also shows radial activity profiles with positive gradients, which become more abundant with increasing stellar mass regardless of environmental density. Our analysis also reveals that the subset of active S0s with negative gradients experiences at all galactocentric radii a systematic reduction in its median activity level with stellar mass, consistent with expectations for main sequence galaxies. In contrast, passive S0s with positive gradients show the opposite behaviour. Furthermore, systems whose activity is dominated by star formation are structurally rounder than the rest of S0s, while those classified as Seyfert exhibit higher rotational support. The possibility that negative and positive activity gradients in S0s may result from rejuvenation by two distinct types of minor mergers is raised.
\end{abstract}


\section{Introduction}
\label{S:intro}
Astronomers have known of the existence of lenticular galaxies (S0s) almost as long as they have known that there are other galaxies besides the Milky Way \citep{Hubble1926}. While it is widely accepted that most spiral galaxies (Sps) emerge from the collapse of primordial vast clouds of gas and dust and subsequent gas-rich mergers that took place in the early Universe \citep[but see][]{Hammer+09,Graham+2023}, and that most elliptical galaxies (Es) have formed from the (often multiple) collisions and mergers of similarly sized galaxies, the formation channel(s) of present-day S0s is (are) still subject to debate. The reason is that, unlike the other two major morphological classes mostly found in specific environments -- Sps in the field and Es around clusters and groups of galaxies -- S0s are relatively abundant in both low- and high-density regions, which points to a complex evolutionary history that may include various pathways. 

Numerical simulations of galaxy interactions show that, in the field or in small groups, S0 galaxies can evolve from mergers, either minor, involving pure discs with small companions, or major between pairs of gas-rich, star-forming (SF) Sps \citep{Querejeta+2015,Tapia+2017,Eliche-Moral+2018}, a scenario that is supported by observations such as those analysed in \citet{Maschmann+2020}. In contrast, physical processes of a complete different nature involving hydrodynamic interactions between the interstellar and intergalactic media \citetext{abbreviated ISM and IGM, respectively; \citealt{Gunn&Gott1972,Nulsen1982}}, probably assisted by other mechanisms such as galaxy harassment or strangulation \citep{Larson+1980,Moore+1999}, are expected to transform Sps into S0s mainly within virialised galaxy clusters \citep[e.g.][]{Solanes+1992,Dressler+1997,Fasano+2000,Quilis+2000,Crowl+2008,Poggianti+2009,DOnofrio+2015}.

It is also becoming increasingly clear that star formation in early-type galaxies (ETGs) -- especially among S0s -- is not a stand-alone event. There is, for example, the study by \citet{Kaviraj+2007} which, in a sample of $\sim 2100$ ETGs that combined UV and optical photometry, found traces of recent star formation in around 30 per cent of the galaxies. Likewise, \citet{Salim+2012}, starting from a small sample of galaxies in the optical red sequence with a strong UV excess but an inactive optical spectrum, determined that galaxy-scale star formation in ETGs is almost exclusively an S0 phenomenon, present in approximately 20 per cent of these objects but practically absent in Es, therefore highlighting the importance of having a disc component to enable extended star formation \citep[see also][]{Mendez-Abreu+2019}.

The study of the origin of lenticular galaxies has recently been boosted by the incorporation of extensive spectroscopic data, both integrated and spatially resolved. Works based on this type of metric have revealed that the S0s show a significant variance in their structural, kinematic and dynamic properties pointing to a bimodality in this population \citep[e.g.][]{Xiao+2016,mckelvie+2018,HDS+2020,Deeley+2020,Rathore+2022,Coccato+2022}, which is problematic to explain by appealing to a single formation channel for these objects. This more than likely multiplicity of evolutionary pathways has also been confirmed by the diverse formative histories associated with galaxies identified with present-day S0s in the newest, state-of-the-art cosmological hydrodynamic simulations that seek to replicate the data provided by large-scale observational studies \citep[e.g.][]{Deeley+2021}.

This paper is the third in a series dedicated to investigate S0 galaxies within the local Universe ($z\lesssim 0.1$) from the wealth of information contained in their optical spectra. These works aim to increase understanding of the physical properties of the members of this morphological class that connects the two ends of the Hubble sequence, with the ultimate goal of providing added insight into the possible formation channels followed by these objects. In our first study \citetext{\citeauthor{Tous+2020}\ \citeyear{Tous+2020}; hereafter \citetalias{Tous+2020}}, we identified three classes of present-day S0s according to their level of activity\footnote{In all papers that conform to this investigation, activity must be understood in a broad sense that includes ongoing star formation, black hole accretion, and/or any other internal source of ionization of the IGM.}  from a sample of $\sim 68,\!000$ of these systems. Our classification is based on the clustering shown by these galaxies in the plane defined by the first two eigenvectors derived from the Principal Component Analysis (PCA) of their integrated optical spectra included in the twelve Data Release of the Sloan Digital Sky Survey \citep[SDSS-DR12;][]{Alam+15}. The projections of the individual S0 spectra onto this 2D subspace are distributed in two clearly delineated regions: the most populated one, which we call the passive sequence (PS), is a compact strip that contains systems conforming to the traditional view of these galaxies as inactive objects, while about a quarter of the local S0 population resides in an extended active cloud (AC), formed by systems with significant nebular emission and high specific star formation rates (SSFR) $\sim 10^{-10.3}\;\mbox{yr}^{-1}$, similar in some cases to those observed in late Sps. Both areas of the diagram are separated by a narrow dividing zone that constitutes the transition region (TR), inhabited by a small fraction of lenticular galaxies with intermediate spectral properties \citetext{see also Section~\ref{SS:dimensionality_reduction} in this work}. 

In a second paper \citep{JimP+22}, we concentrated our efforts on comparing the outcomes of our PCA spectral classifier of S0 galaxies with those delivered by well-known taxonomic paradigms of activity, such as the BPT \citep{Baldwin+1981} and WHAN \citep{Cid+2010} diagrams, as well as other conventional activity diagnostics defined in the radio, mid-infrared, and X-ray domains. In this work, we demonstrated that our definition of the PCA spectral classes correlates very strongly with the distance of galaxies to the ridge line of the PS in the PC1--PC2 subspace 
\begin{equation}
    \mbox{PC2} = -1.162 \cdot \mbox{PC1} - 2.932\;,
    \label{eq:ps}
\end{equation}
which is given by the equation:
\begin{equation}
    \Delta\mbox{PS} = \log\left[2.913+0.758\cdot\mbox{PC1}+0.652\cdot\mbox{PC2}\right]\;,
    \label{eq:DPS}
\end{equation}
thus reducing in practice the PCA taxonomy to a one-dimensional classifier of activity. In addition to making it possible the compression of the optical spectra of S0 galaxies into a single parameter, this distance is an excellent proxy for the equivalent width (EW) of the \Ha\ emission, to the point that the constant values of \DPS\ used to set the PS/TR and TR/AC dividers, $0.23$ and $0.44$, respectively (see also Fig.~\ref{fig:ps_pc1_pc2}), bear a one-to-one correspondence with the $3$ and $6\,\angs$ horizontal demarcations of the $\mbox{EW}($\Ha$)$ in the WHAN diagram. With regard to the sources of activity, we showed that the AC--S0 subpopulation contains a mixture of SF galaxies (which are the majority) and mostly low-power (i.e.\ Seyfert 2) AGNs, while LINERs and retired galaxies make up, respectively, the bulk of the TR and PS spectral classes. We also found evidence that most of the nebular emission of Seyfert and LINER S0 systems detected in radio and X-ray is not driven by star birth, and that the dominant ionizing radiation for a good number of LINERs could come from post-AGB stars. All these findings inevitably raise questions about the mechanisms that trigger the episodes of activity in lenticular galaxies.

Recently, in \citet{Tous+2023}, we have made our first attempt to answering these queries, aided by integral field spectroscopic data from 532 S0 galaxies drawn from the first data release of the Mapping Nearby Galaxies at Apache Point Observatory \citetext{MaNGA; \citealt{Bundy+2015}} survey. In such work, we have put the focus on aspects exclusively related to star-formation activity and, in particular, to the presence in these galaxies of SF rings in the EW(\Ha). Assessment of these ringed structures indicates that they are relatively abundant ($\sim 30$ per cent on average) in fully-formed S0s, but with a frequency that sharply increases with the mass of the hosts, that they are twice more frequent among the members of the PS class than in AC--S0s, that they likely feed on residual gas from the discs, and that their formation is not influenced by the environment. These results suggest, in good agreement with cosmological simulations, that the formation of these \Ha\ rings is possibly associated with the capture by the S0s of tiny dwarf satellites that closely orbit them.

In this new paper, we aim to characterize in detail the radial structure of activity defined by the baryonic content of lenticular galaxies with the expectation of providing further clues regarding the evolutionary history of this morphological family. For this task, we shall rely again on spatially resolved spectroscopic observations, in this case on a sample of 1072 present-day S0s drawn from the final data release of the SDSS-IV MaNGA survey. In Section~\ref{S:data}, we present these spectral data and the ancillary properties that are used throughout this study, while the processing of the MaNGA data cubes is described in Section~\ref{S:data_cube_processing}. After identifying the main sources of activity in our S0 sample (Section~\ref{S:activity_profiles}), we determine in Section~\ref{S:profiles_vs_properties} the typical shapes of the radial activity profiles of the different spectral classes of these systems and examine their relationship with a number of physical magnitudes that are traditionally used to diagnose galaxy evolution. Finally, in Section~\ref{S:discussion}, we review the agreement of our results regarding the activity morphology of nearby S0 galaxies with predictions from the main evolutionary channels of these objects proposed in the literature. Section~\ref{S:summary} provides a summary of this work. The manuscript also includes an Appendix~\ref{A:morphological_contamination} where we test the robustness of our results against morphological contamination.

The values of all cosmology-dependent quantities used in the paper have been calculated by assuming a  concordant flat ${\rm \Lambda}$CDM universe with parameters $H_0 = 70\;{\rm km\;s^{-1}\;Mpc^{-1}}$ and ${\rm \Omega_m} = 1 - {\rm \Omega_\Lambda} = 0.3$.

\section{The data}
\label{S:data}
We use the data cubes from the final data release (DR17, \citealt{Abdurro+2022}) of the Mapping Nearby Galaxies at Apache Point Observatory (MaNGA) survey \citep{Bundy+2015}, which is part of the Sloan Digital Sky Survey IV (SDSS, \citealt {Blanton+2017}). This survey provides spatially resolved spectroscopic information up to $1.5$ effective radii (\Reff) in the $r$ band\footnote{Although $\sim 1/3$ of the MaNGA galaxies are covered up to $2.5$~\!\Reff, in this work we restrict the analysis to the information contained within $1.5$~\!\Reff\ to maximise sample size.} for a sample of $\sim 10000$ nearby ($\tilde{z}\sim 0.03$) galaxies observed in the visible ($3500 - 10000$ \AA) with a spectral resolution of $R=\lambda/\mbox{FWHM}\sim 2000$ \citep{Sme+13}, a spatial resolution of $\sim 1\,$kpc at the median redshift $\tilde{z}$ of the sample. We also use the 2D maps of physical parameters from both the continuum and the ionized emission lines produced by the Data Analysis Pipeline (DAP) of MaNGA \citep{Westfall+2019} through the HYB10-MILESHC-MASTARHC2 analysis approach. We extract from these maps the in-plane disc radial coordinate of the $M$ spaxels of each galaxy in the dimensionless form $R^\prime \equiv R/$\Reff, as well as the line-of-sight stellar velocity and velocity dispersion. These data are accessed and manipulated with the aid of the \textsc{Marvin}\footnote{\url{https://sdss-marvin.readthedocs.io/en/stable/}} toolkit \citep{Che+19}.

Lenticular galaxies are selected from the MaNGA Deep Learning Morphological value added catalogue (MDLM-VAC-DR17) of \citet{HDS+2022}. Following this work, our S0 sample is defined from the numerical Hubble stage, $T$, the probability that a galaxy is of late-type, $P_{\mathrm{LTG}}$, and the probability that an early-type looking object is truly an S0, $P_{\mathrm{S0}}$, by applying the conditions $T < 0.5$, $P_{\mathrm{LTG}} < 0.5$, and $P_{\mathrm{S0}} > 0.5$. These restrictions are fulfilled by $1283$ galaxies, which reduce to $1072$ once we remove a few problematic cases flagged by the MaNGA Data Reduction Pipeline (DRP), as well as those galaxies for which we detect, after a careful visual inspection of the SDSS colour images, a hint of spiral structure, and or contaminating foreground stars, companion galaxies, or both within the MaNGA's field of view. 

The global activity status of each one of the S0s, which we define by their spectral class, is determined from our own PCA-based classification (see also Section~\ref{SS:dimensionality_reduction}). For most galaxies, we are able to directly use the class assigned in \citetalias{Tous+2020} from their distance to the ridge of the passive sequence (\DPS; see eqs.~(\ref{eq:ps}) and (\ref{eq:DPS})) measured in the latent space produced by the first two principal components (PC) of the single-fibre SDSS spectra of the present-day S0 galaxy population. However, there are 128 galaxies in our current sample that are not included in the SDSS Legacy Survey or that, although they are part of it, have not received a spectral classification because of problems with their single-fibre spectra (e.g.\ an excessive number of masked regions). For these objects, we first synthesize their integrated spectra by emulating the three-arcsecond aperture of the SDSS fibres from the MaNGA data cubes and then project the synthetic spectra onto the PC1--PC2 subspace to determine the spectral class. We have verified that the results of this work do not change if these 128 galaxies are excluded. By proceeding in this way, we find that our MaNGA S0 sample contains a total of $781$ galaxies of the PS spectral class, $199$ of the AC class, and $92$ TR objects (see Table~\ref{tab:sample_sizes}).

The spectral information is extended with several physical magnitudes that are relevant for the aims of this work. Thus, we include for a great deal of the MaNGA S0s the values of their local environmental density estimated in \citetalias{Tous+2020}. They consist on the percentile associated to this property when it is calculated from the average projected distance between a galaxy and its first five neighbours with recessional velocities less than 1000 km$\,\mbox{s}^{-1}$ from it in the deepest volume-limited region that can be defined for the central object in the $z < 0.1$ subset of the SDSS Legacy Survey \citetext{see eqs. (1) and (2) in this work}. We also retrieve, for most galaxies, their global stellar mass, $M_\ast$, and star formation rate (SFR) by crossmatching our dataset with the GALEX-SDSS-WISE Legacy Catalog 2 (GSWLC-2; \citealt{SBL18}), while the luminosity- and mass-weighted mean ages of the stellar populations are obtained from the MaNGA Pipe3D value added catalogue for those galaxies with a null QCFLAG \citep{Sanchez+2016a, Sanchez+2016b, Sanchez+2022}. The best estimates of the apparent axis ratio $b/a$ in the $r$ band, taken from the latest version of the PyMorph photometric catalogue \citep{HDS+2022}, are used to determine the ellipticity of each galaxy, defined as $\varepsilon = 1 - b/a$. Finally, we infer for all our galaxies their observed projected stellar angular momentum per unit mass, $\lRe$, by adopting the definition introduced by \citet{Emsellem+2007}:
\begin{equation}
    \lRe = \dfrac{\sum\limits_{j=1}^{M_\mathrm{e}}{R_j F_j |V_j|}}{\sum\limits_{j=1}^{M_\mathrm{e}} R_j F_j \sqrt{V_j^2 + \sigma_j^2}}\;,
    \label{eq:angular_momentum}
\end{equation}
where $F$, $V$, and $\sigma$ are, respectively, the flux, and the stellar rotational velocity and velocity dispersion corrected for instrumental resolution of the $j$th spaxel, and the sum runs over the $M_\mathrm{e}$ spaxels within elliptical isophotes of mean radius $R=a\sqrt{1-\varepsilon}$ out to $1\,R_{\rm e}$.

\section{Data cube processing}
\label{S:data_cube_processing}
Next, we proceed to divide the MaNGA data cubes into annuli by grouping the individual spectra of the spaxels into ten circularized radial bins defined in terms of the dimensionless variable $R^\prime$ (assigned to each spaxel by the DAP) using steps of $0.15 R^\prime$ within the interval $[0,1.5]$. 

\subsection{Build-up of the radial composite spectra}
\label{SS:stacking}
In order to stack the spectral information of the spaxels included within a radial bin, we follow a method that is an adaptation of the one used in \citeauthor{Navo+2019} (\citeyear{Navo+2019}; see also \citealt{Mas+2017}). To begin with, we exclude from the stacking any spaxel with $\mbox{S/N}<5$. Next, for those spaxels above this limit, we mask any problematic flux element flagged by the MaNGA's DRP, as well as those flux elements in wavelength ranges that are usually affected by sky lines. Nonetheless, we have observed that unrealistic flux values sometimes escape detection by these filters and end up contributing to the final composite spectrum. Although only about one per cent of the processed spaxels contain one or more glitches, they can greatly influence the PCA. To limit as much as possible the presence of these false emission lines, we apply the following procedure to identify and mask them:
\begin{enumerate}
    \item  For a given spectrum, the maximum flux value is determined every 50 consecutive elements of the flux vector.

    \item From the distribution of maximum flux values, a threshold is established above of which any flux value will be considered as a potential glitch candidate. The threshold is defined as the sum of the third quartile and 10 times the interquartile range (IQR) of the distribution of maximum flux values. The arbitrary factor multiplying the IQR has been chosen to make the threshold lay above most of the actual strong emission lines.
    
    \item To ensure that any real emission line is not being masked, the candidate glitches are cross-checked against an extensive list of known emission lines\footnote{Retrieved from \url{http://astronomy.nmsu.edu/drewski/tableofemissionlines.html} conveniently converted from air to vacuum wavelengths as stated in \url{https://www.sdss4.org/dr17/spectro/spectro_basics/}.}. 
    
    \item Candidate glitches are considered true glitches either if they are located at a distance larger than $3$ \AA\ from any of the known emission lines, or if they are near a known emission line but have a maximum flux value which is at least $1.2$ times larger than the highest of the peak value of the fluxes of the lines [\ion{O}{III}]$\,\lambda\, 4960$, [\ion{O}{III}]$\,\lambda\,5008$, \ion{H}{$\beta$}, \ion{H}{$\alpha$}, or [\ion{N}{II}]$\,\lambda\,6585$ of the same spectrum.
\end{enumerate}
Any spaxel that after this second filtering step has more than $10$ per cent of its flux elements masked is also discarded. Retained spectra are then corrected for Galactic dust reddening using a standard \citet{Fitzpatrick+1999} dust extinction model. To do so, we determine the total extinction at $V$, $A_{V}$, by renormalizing the extinction coefficient of each object in the $g$ band listed in the NASA-Sloan Atlas catalog \citep{Blanton+2011} by the factor $R_{V}/R_{g}=0.82$. Individual spectra are subsequently shifted to the laboratory rest-frame through the following expression that removes the recessional velocity of each galaxy and the mean rotation speed from each spaxel: $\log \lambda_{\mathrm r} = \log \lambda_{\mathrm o} - \log(1 + z) - \log(1 + v/c)$, where $\lambda_{\mathrm r}$ and $\lambda_{\mathrm o}$ are, respectively, the rest-frame and observed wavelengths, while $v$ is the rotational velocity of the spaxel. The flux and associated error vectors are re-binned by interpolating into new intervals with the same constant logarithmic spacing of $0.0001$ of the original spectra.

Then, we determine the normalization factor for each individual spectrum, which for a given galaxy is defined as
\begin{equation}
	n_j = \sum\limits_{i=1}^{N^\mathrm{c}_j} \frac{f_{ij}}{N^\mathrm{c}_j}\;,
	\label{eq:norm}
\end{equation}
where $f_{ij}$ is the $i$th flux element or wavelength bin of the $j$th spaxel of the galaxy, $f_j$, and the summation runs over the $N^\mathrm{c}_j$ flux elements of this spaxel included in the four rest-frame wavelength intervals $4200\:$--$\:4300$, $4600\:$--$\:4800$, $5400\:$--$\:5500$, and $5600\:$--$\:5800$ \AA\ that are good representatives of the continuum \citep{Dobos}. Besides, to weight each spaxel in a manner that penalizes the noisiest ones, we also compute the S/N of the continuum,
\begin{equation}
    w_j = n_j \left( \sum\limits_{i=1}^{N^\mathrm{c}_j} \frac{e_{ij}^2} {N^\mathrm{c}_j} \right)^{-1/2}\;,
    \label{eq:weight}
\end{equation}
where $e_{ij}$ is the error associated with the $i$th flux element of the spectrum of the $j$th spaxel and the summation runs over the same wavelength intervals defined for equation~(\ref{eq:norm}). Note that $N^\mathrm{c}_j$ may differ from one individual spectrum to another depending on the number of masked flux elements that fall on the adopted continuum regions. Spaxel spectra with more than $20$ per cent of masked flux values within these intervals are discarded too. 

Finally, we derive the composite spectrum for the $k$th radial bin of a galaxy by means of the equation
\begin{equation}
    \bar{f_k} = \sum\limits_{j = 1}^{M_k} w_j \dfrac{f_{j}}{ n_j}\left(\sum\limits_{j = 1}^{M_k} w_j\right)^{-1}\;,
    \label{eq:composite_flux}
\end{equation}
where the summation runs over the $M_k$ spaxels included in this radial bin. We restrict this calculation to radial bins for which at least $20$ per cent of their spaxels have spectra that meet the aforementioned quality conditions.

After applying this stacking process, the total spectral information of all the galaxies in our sample is stored in a single large matrix that contains, for each galaxy, $k \leq 10$ rows or elements of the radial composite spectra and $\sim 3300$ columns, one per each flux element of the radial flux arrays. 

\subsection{Dimensionality reduction through PCA}
\label{SS:dimensionality_reduction}
To facilitate the interpretation of the information encoded in the radial spectra of galaxies, we now proceed to radically reduce their dimensionality but keeping most of the variance contained in the original 2D arrays. As was done in the two previous papers of this series \citetext{\citetalias{Tous+2020}, \citealt{JimP+22}}, as well as in \citet{Tous+2023}, to accomplish this we apply a multivariate analysis statistical technique known as PCA. In short, this information compression method transforms the original flux arrays by rotating the $N$-dimensional space defined by their flux elements into a new orthogonal coordinate system whose eigenvectors are both oriented along the directions in which the variance of the data is maximized and ranked in decreasing order according to the fraction of the total variance they encapsulate. This allows one to replace the initial highly-dimensional space with a low-dimension space defined by a few first eigenvectors or principal components using the tuples of scalars that result from projecting the original spectra onto such eigenvectors\footnote{By abuse of language, the term principal component is often used interchangeably both for the eigenvectors and for the scalar coefficients associated with the projections of the data on them.} with a minimal loss of information \citetext{see \citetalias{Tous+2020} for more information on this technique}.

\begin{figure}
    \centering
	\includegraphics[width=\columnwidth]{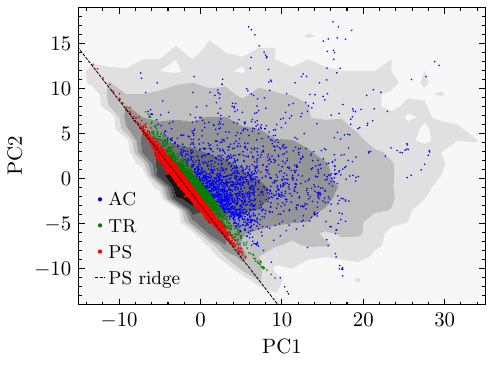}
    \caption{PC1 and PC2 components of the radial composite spectra of the 1072 lenticular galaxies included in our MANGA S0 sample  projected onto the eigenbasis defined in \citetalias{Tous+2020} from a sample of more than $68{,}000$ SDSS spectra of present-day ($z\lesssim 0.1$) S0s. The greyscale contours represent equally spaced logarithmic densities of the projections inferred in that work. The radial elements of the composite spectra that belong to the PS ($\mbox{\DPS}<0.23$), the TR ($0.23\leq\mbox{\DPS}<0.44$) and the AC ($\mbox{\DPS}\geq0.44$) are drawn, respectively, as red, green and blue dots. The black dashed line shows the ridge of the PS given by equation~(\ref{eq:ps}).}
    \label{fig:ps_pc1_pc2}
\end{figure}

As in \citetalias{Tous+2020}, we restrict the current analysis to the first two principal components, since they already offer a very complete description of the variance of the spectra of present-day S0 galaxies. Unsurprisingly, the representation we obtain of the MaNGA's radial composite spectra in the PC1--PC2 subspace defined by \citetalias{Tous+2020} bears a great similarity to that inferred in this previous work from a much larger sample of single-fibre SDSS spectra of nearby S0s ($z\lesssim 0.1$). As shown in Fig.~\ref{fig:ps_pc1_pc2}, we have, on one hand, a narrow and densely populated diagonal band delineated by the clustering of the projections of fully passive spectra from objects that belong to the PS class. This region is characterized by a sharp linear ridge that crosses the subspace from small/large values of PC1/PC2 to larger/smaller values, and that sets the zero point that we use for the spectral classification of the S0s \citep[see also][]{JimP+22}. On the other hand, a much more dispersed group of data can be observed to the right of the passive sequence, which extends over a large area of the subspace. This cloud of points encompasses all the radial composite spectra that exhibit emission lines -- the relevance of which progressively increases as they move away from the PS ridge -- and which are all part of the AC class. Between these two main areas of the diagram, it is possible to identify, in a very narrow strip that runs parallel to the PS ridge, spectra of the TR class, with intermediate characteristics from those of the two previous groups. 

Our PCA-based taxonomy of galaxy spectra has several advantages compared to more traditional procedures, among them the fact that it is a versatile methodology applicable to any galaxy, irrespective of its activity level. In contrast, common diagnostics in the literature like the BPT and WHAN diagrams, rely on the flux (ratios) of a few specific emission lines (\Ha, \Hb, \Nii, \Oiii, $\ldots$), making them less effective for classifying spectra with low signal-to-noise ratios or lacking these lines, as in classical ETGs. Since our approach utilizes the entire optical spectrum, it also eliminates the need for continuum subtraction, for which there is no standardised procedure. Furthermore, our spectral classification stems from purely mathematical arguments derived from the observed distribution of the principal components in the latent space, avoiding the necessity for a priori assumptions about the physics of galaxies to define model-based or theoretically motivated dividers. And while the quantities derived from PCA are not necessarily physical parameters, this method allows us to explore whether it is possible to characterize gas ionization in galaxies more optimally than through the physical properties used in emission-line-based diagnostics. As discussed in Section~\ref{S:activity_profiles}, our analysis yields indicators that are effective proxies for well-known specific sources of gas ionization.

In comparing the outcomes from the PCA of the MaNGA and SDSS spectra of S0 galaxies, it is important to remark that the direction of the PS ridge (equation~(\ref{eq:ps}) and dashed line in Fig.~\ref{fig:ps_pc1_pc2}) has not changed. Although the instrument used in obtaining the spectra of both samples is the same, as it is the way of presenting the data to the PCA, the differences in the spatial coverage of the observations, in the dynamic range of the targeted objects, and even in the methodology and strictness adopted in the morphological identification of the S0s, will necessarily translate into a different covariance matrix and result in the reorientation of the orthogonal base of eigenvectors derived from the PCA decomposition of the MaNGA data. However, the contribution to the total variance of these factors is expected to be quite moderate and have a minimum impact on the firsts principal components. In fact, we have verified that the reorientation of the MaNGA's eigenbasis hardly affects the first two principal components, and therefore neither does it affect the projections of its spectra on said axes, the changes only beginning to be noticeable from the third component. This invariance is also transferred to the PS ridge lines deduced from the MaNGA and SDSS spectra which, calculated in the same PC1--PC2 plane, have orientations that differ by less than one degree and intercepts that differ by just one tenth. Thus, the use in this work of the same PC1--PC2 subspace, PS ridge line, and \DPS\ demarcations defined in \citet{JimP+22} to classify the radial composite spectra of the MaNGA S0 sample is fully justified.
% Fig. 1

\subsection{Vectorization of the projected spectra}
\label{SS:vectorization}

\begin{figure*}
    \centering
    \begin{subfigure}{0.50\textwidth}
        \includegraphics[width=\linewidth]{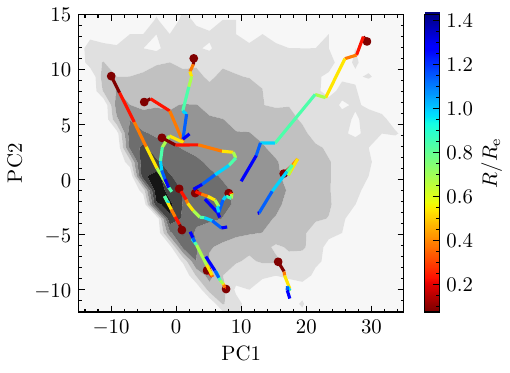}
    \end{subfigure}
    \begin{subfigure}{0.49\textwidth}
        \includegraphics[width=\linewidth]{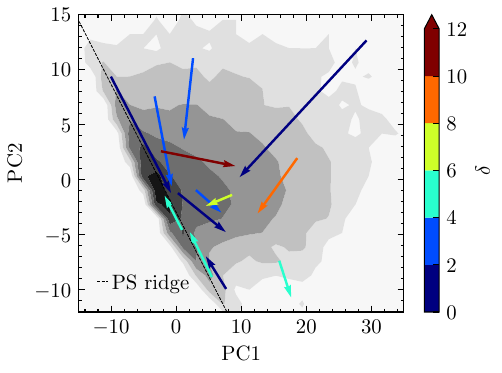}
    \end{subfigure} 
    \caption{\emph{Left:} radial profiles of a small random subset of the MaNGA S0 sample in the subspace of the first two PCs defined in \citetalias{Tous+2020} for nearby lenticulars. 
    The composite spectra in each radial bin are coloured according to the galactocentric distance of the bin scaled by the effective radius of the galaxies, with brown dots indicating the position of the central (innermost) bins. \emph{Right:} vectorization of the spectral profiles in the left panel. The colours show the value of $\delta$, one of the two parameters used to evaluate its feasibility (see text). The brown arrow identifies one profile with $\delta\geq 10$, unsuitable for vectorization. The black dotted line shows the PS ridge. In both panels, the greyscale contours are the same as Fig.~\ref{fig:ps_pc1_pc2}.}
    \label{fig:vectorization}
\end{figure*}

The fact that a large part ($\sim 87$ per cent) of the sample variance of the radial spectral profiles is captured by their first two principal components allows us to carry out the analysis of the MaNGA's spectra using their set of projections in the more convenient PC1--PC2 latent space\footnote{The first five eigenvectors, including the two that define this PC1--PC2 latent space, are publicly available in \citet{JimP+22_table}.}, where such profiles are much easier to handle and interpret. Moreover, the approximate linear behaviour shown by most of these strings of PC1 and PC2 components makes it possible to further streamline their analysis without missing essential information. This additional simplification consists in approximating the PC1--PC2 representations of the radial profiles with best-fitting straight lines, which are then converted into vectors using the projections of the PC1--PC2 coordinates of the innermost and outermost radial bins of the spectra on such lines to set, respectively, the tail and tip of the arrows. To objectively identify the few profiles that do not admit a good vectorization (see, e.g., the profile in Fig.~\ref{fig:vectorization} associated with a brown vector in the right-hand panel), we have computed two ad-hoc parameters that measure the goodness of such representation. The first one is called the disorder, since it determines the extent to which the projections of the PC1 and PC2 coordinates of the radial bins of a spectrum into the fitted straight line follow the right sequence of radial distances, 
\begin{equation}
\delta = \left(\sum_{k=0}^9 (p_k - k)^2\right)^{-1/2}\;,\ \ \ \ \ \ \ \ \ \ \ \{p_k \in \mathbb{Z}\ |\ 0 \leq p_k \leq 9\}\;,
\label{eq:disorder}
\end{equation}
where $p_k$ is the position of the $k$th radial bin within the vector components. The second parameter, which we call the entanglement, provides a measure of the degree of linearity of the radial spectral profiles in the PC1--PC2 plane from the ratio between the second and the first singular values, i.e.\
\begin{equation}
    \epsilon = \dfrac{\sigma_2}{\sigma_1}\; ,
    \label{eq:entanglement}
\end{equation}
of the $10\times 2$ matrix, $M$, whose rows are the PC1 and PC2 coordinates of the radial bins of a profile. The singular values are found by diagonalizing  $M^T M$: 
\begin{equation}
M^T M = V(\Sigma^T \Sigma)V^T\; ,
\label{eq:svd_diagonal}
\end{equation}
where $V$ is a $2\times 2$ matrix whose columns are the eigenvectors of $M^T M$ and the non-zero elements of the $10\times 2$ rectangular diagonal matrix $\Sigma$ are the square roots of the non-zero eigenvalues, $\lambda_i$, of $M^T M$, so $\sigma_i = \sqrt{\lambda_i}$ with $i = 1, 2$. After some experimentation, we have decided to restrict the vectorization to those profiles that have either $\delta < 7$ for any $\epsilon$ or $\delta < 10$ and $\epsilon < 0.06$. About 90 per cent of the galaxies in our sample show radial spectral profiles that are linear according to these criteria. The results of our analysis do not vary if other sensible combinations of upper bounds for these two parameters are considered instead.

% Table 1
\begin{table*}
\begin{threeparttable}
\centering
    \caption{Number and fraction of S0 galaxies included in the different subsamples considered in this work.}
    \label{tab:sample_sizes}
    \begin{tabular}{crrrrr}
    \hline
    \multicolumn{1}{c}{Spectral class}    & \multicolumn{1}{c}{All$^a$}         & \multicolumn{1}{c}{W.\ linear profiles$^b$} & \multicolumn{1}{c}{Outside in$^b$} & \multicolumn{1}{c}{Inside out$^b$} & \multicolumn{1}{c}{Flat$^b$}      \\ \hline
    Passive sequence  & 781 (0.73)    & 688 (0.73)\ \ \ \ & 31 (0.05)\ \ \  & 177 (0.26)\ \ \ & 480 (0.70) \\
    Active cloud      & 199 (0.19)    & 188 (0.83)\ \ \ \ & 148 (0.79)\ \ \ & 30 (0.16)\ \ \ & 10 (0.05)  \\
    Transition region & 92 (0.09)     & 84 (0.72)\ \ \ \ & 32 (0.38)\ \ \   & 27 (0.32)\ \ \ & 25 (0.30)  \\ 
    \hline    
    All               & 1072 (1.0)     & 960 (0.90)\ \ \ \  & 211 (0.22)\ \ \   & 234 (0.24)\ \ \ & 515 (0.54) \\ \hline
    \end{tabular}
    \begin{tablenotes}\footnotesize
    \item[] In parentheses are the fractions of galaxies ($^a$) in each spectral class and ($^b$) with linear profiles in the same spectral class.
    \end{tablenotes}
\end{threeparttable}
\end{table*}

The vectorization of the composite radial spectra facilitates their division into separate categories according to their orientation relative to the main ridge of the PS, i.e.\ according to the sign of the radial gradient of their activity. The latter is given by the polar angle, $\theta$, of the associated vectors measured counterclockwise from the ridge line. We define three orientation tiers: 
\begin{itemize}
\item galaxies with $5^\circ < \theta <       175^\circ$, which correspond to objects whose activity grows inside out (IO) up to $\sim 1.5 R_{\mathrm e}$, i.e.\ that show a positive radial gradient of activity within this galactocentric radius; 
\item galaxies with $185^\circ < \theta < 355^\circ$, for which the gradient of activity is negative as it grows outside in (OI)\footnote{Note that these designations may not necessarily indicate inside-out or outside-in quenching, a common terminology used to state that the cessation of star formation progresses from the centre to large radii or vice versa.}; and 
\item galaxies with $\theta \in [-5^\circ, 5^\circ] \cup [175^\circ, 185^\circ]$, whose radial activity profiles run essentially parallel to the PS ridge and that, therefore, exhibit a null or flat (F) activity gradient. 
\end{itemize}

As shown in Table~\ref{tab:sample_sizes}, the majority ($70$ per cent) of the PS lenticulars in our sample whose composite radial spectra support vectorization show F-type profiles, while $26$ per cent of such systems have IO activity gradients and only a few objects ($5$ per cent) OI gradients. In the case of the AC class, galaxies with OI gradients clearly dominate ($79$ per cent), while those with F profiles are in the minority ($5$ per cent). And for TR class objects, the signs of the radial activity gradients are approximately evenly distributed, although with a slight prevalence of those of type OI. Vice versa, by reading the table from the point of view of the gradients, one can see that $70$ per cent of the galaxies with OI activity profiles belong to the spectral class AC, while up to $93$ per cent of those with F profiles are PS systems and only less than $2$ per cent objects of the AC class. Finally, $\sim 3$ out of every 4 positive activity gradients (i.e.\ IO profiles) occur in PS--S0s.
% Fig. 2

We note that we have chosen not to impose cuts on inclination or apparent size of galaxies relative to the point spread function (PSF) within our S0 sample, after verifying that including objects with extreme values of these observational parameters ($i > 70^\circ$, $\sigma_{\mathrm{PSF}}/$\Reff$ > 0.5$) does not bias our analysis. While applying these additional selection criteria would result in the exclusion of a non-negligible number of galaxies ($\sim 8$ and 16 per cent, respectively), the fact that our approach relies on both the sign of activity gradients rather than their specific numerical values, along with the strict angular boundaries we apply to identified a gradient as flat, effectively mitigates the flattening in the radial profiles that oversampled and highly inclined objects can induce. In any case, a thorough examination of the outcomes both before and after implementing the aforementioned cuts to the data, using the same battery of empirical tests outlined in Appendix~\ref{A:morphological_contamination}, confirms that retaining such galaxies in the sample does not compromise the main findings and conclusions of this study.

\section{Activity profiles of S0 galaxies}
\label{S:activity_profiles}
\begin{figure}
    \centering
	\includegraphics[width=\columnwidth]{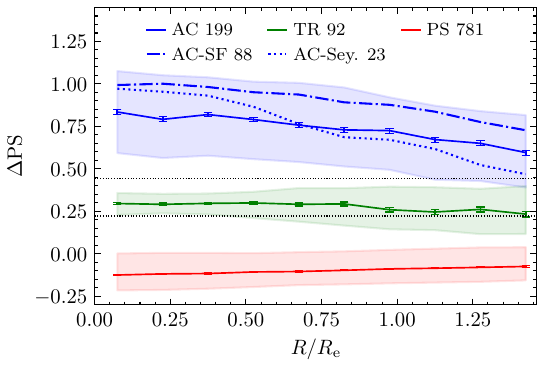}
    \caption{Median radial \DPS\ profiles of 1072 present-day S0s observed by MaNGA segregated with respect to spectral class into passive sequence (PS; red), transition region (TR; green), and active cloud (AC; blue) systems. Error bars correspond to the $1\sigma$ scatter of the median. Shaded regions show the interquartile range of the profile distribution within each radial bin. For the AC subset, we also show the spectral profiles of galaxies classified by \citet{Thomas+2013} as star forming (SF, blue dash-dotted line) and Seyfert (Sey., blue dotted line). The figures in the legend report the total number of objects considered in deriving the statistics for each subset. The horizontal dotted lines, located at $0.23$ and $0.44$ in \DPS, indicate the boundaries of the TR as defined in~\citet{JimP+22}.}
    \label{fig:dps_profiles}
\end{figure}

In this section, we use the PC1 and PC2 coordinates of the radial spectral profiles of our S0 galaxies to derive their corresponding \DPS\ profiles in order to infer how activity modulates with galactocentric distance in these objects. Fig.~\ref{fig:dps_profiles} shows the interquartile range of the distributions of the \DPS\ profiles for each of the three spectral classes of S0s. Leaving aside the shapes of the corresponding median profiles (solid curves), which are perfectly explained by the data listed in Table~\ref{tab:sample_sizes}, a first result that emerges from this graph is that, within the range of galactocentric distances examined by MaNGA, the local estimates of the spectral class of most S0s agree with the global spectral class deduced from their respective 3-arcsec-single-fibre SDSS spectra, regardless of the physical sources of the activity. This is true for nearly $95$ per cent of the PS-class galaxies, about $75$ per cent of the AC--S0s, and $50$ per cent of the TR members, a more than considerable proportion given the narrow range of \DPS\ values that define this latter class. 

Another interesting result shown in Fig.~\ref{fig:dps_profiles} is that the median radial spectral profile of the subset of AC--S0s assembled from galaxies classified as SF in a classical BPT diagram (dot-dashed blue curve) remains at all radii above, not only of the median profile of the AC subset as a whole, but also of that inferred from their counterparts classified as Seyfert (dotted blue curve), including the innermost region. This suggests that the highest values of \DPS\ in present-day S0s will be normally achieved by objects that are actively forming stars (but see next section). However, it must be kept in mind that \DPS\ is a \emph{mathematical} parameter sensitive to any type of nebular ionization source that has its reflection in the optical spectra of galaxies, be it ongoing star formation, accretion into central supermassive black holes (SMBH), photoionizing radiation from post-AGB stars, or wind-driven shocks in the ISM. Therefore, before proceeding further, it is important that we gather all the information that can help us identify the main physical mechanisms contributing to \DPS\ in each of the three spectral classes of lenticular galaxies. 

% Fig. 3

% Fig. 4
\begin{figure*}
    \centering
	\includegraphics[width=0.95\textwidth]{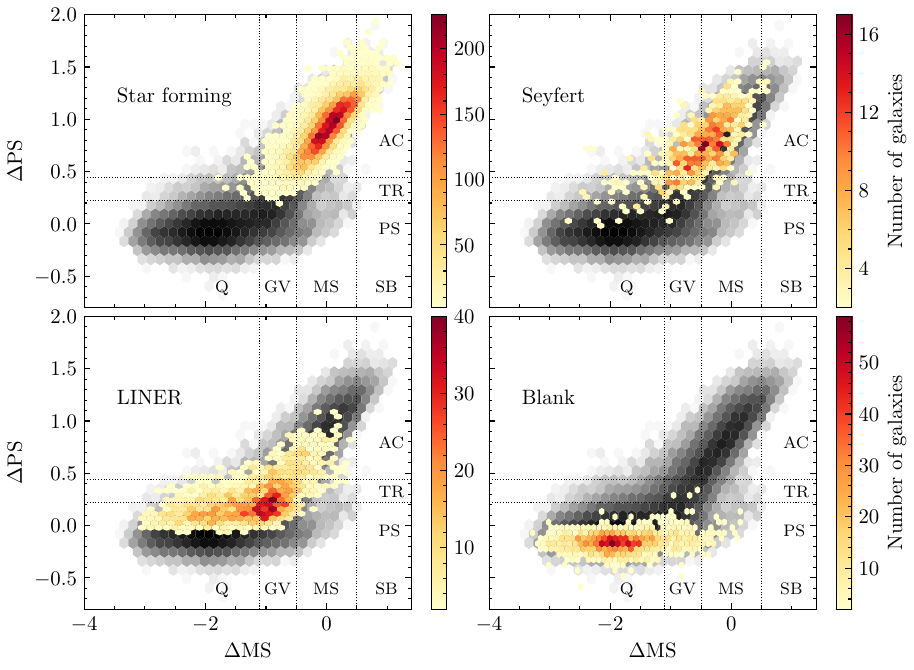}
    \caption{Distance to the ridge of the passive sequence (\DPS) vs distance to the MS of SF galaxies (\DMS) for the sample of $\sim 68,000$ S0s studied in \citetalias{Tous+2020}. The grey-shaded background binned into hexagonal pixels shows the distribution of the full sample, the darker the tones the higher the number density. Overlaid coloured hexagonal pixels show the distributions of those lenticulars from this dataset classified by \citet{Thomas+2013} as star forming (upper-left panel), Seyfert (upper-right), LINER (bottom-left), or Blank, i.e.\ without a BPT \citep{Baldwin+1981} classification due to the lack of emission lines (bottom-right). In the first three panels only objects with an amplitude-over-noise $> 2$ in the four emission lines [\ion{O}{III}]$\,\lambda\,5008$, $\text{H}\,\beta$, $\text{H}\,\alpha$, and [\ion{N}{II}]$\,\lambda\,6585$ have been considered. The horizontal dotted lines show the demarcations of our three spectral classes, from top to bottom, active cloud (AC), transition region (TR), and passive sequence (PS), whereas the vertical dotted lines divide the galaxies, from left to right, into quiescent (Q), green valley (GV), main sequence (MS), and starburst (SB) systems.}
    \label{fig:bpt_dps_dms}
\end{figure*}

With this aim, we show in the background of the four panels of Fig.~\ref{fig:bpt_dps_dms} the bivariate distribution of the \DPS\ values of the sample of single-fibre S0 spectra studied in \citetalias{Tous+2020} as a function of their separation from the MS of SF galaxies, \DMS, using as zero point the ridge line of the star formation peak \citep[cf][]{Renzini+2015}:
\begin{equation}
\mathrm{\Delta MS} = \log\left[\frac{\mathrm{SFR}}{\mbox{M}_\odot\,\mathrm{yr}^{-1}}\right]-0.76\log\left[\frac{M_\ast}{\mbox{M}_\odot}\right] + 7.6\;.
\label{eq:DMS}
\end{equation}
The panels include the \DPS-based demarcations of our three spectral classes, as well as the same \DMS\ dividers used by \citet{Bluck+2020b} for the general population of galaxies, which we use to assign different statuses to the S0s according to the value of this parameter, that is a measure of the SSFR. We have also overlaid on each panel coloured hexagonal cells that show, from left to right and top to bottom, the distributions of the S0s in that sample classified by \citet{Thomas+2013}, using the standard \ion{N}{II}--BPT line-ratio diagnostic, as SF, Seyfert, or LINER, while the fourth panel shows the distribution of those S0s identified as `blank' because their emission lines were unavailable. Unsurprisingly, the latter are objects that in our PCA-based taxonomic paradigm are included in the PS spectral group and that, for the most part, have negative, uncorrelated \DPS\ and \DMS\ values, as well as a quiescent status regarding the star formation. On the other hand, the vast majority of both SF- and Seyfert-type lenticulars are found within the AC class. Unlike the above, in the present case both types of objects show a strong positive linear correlation between their values of \DPS\ and \DMS, especially those belonging to the SF category, which, in agreement with the work of \citet{JimP+22}, are also by far the most abundant within the AC subpopulation. However, it can be observed that when the AGN activity dominates, star formation tends to be somewhat less important, suggesting negative feedback. In addition, while the maxima of the distributions of both BPT types lie in the MS, Fig.~\ref{fig:bpt_dps_dms} shows that the most active of the AC--S0s are SF objects that extend into the starburst (SB) region, while the least active of these systems are distributed along the green valley (GV), even penetrating, more clearly when it comes to Seyfert-type galaxies, the quiescent (Q) zone. We find up to nine MaNGA S0s located in this sparsely populated Q--AC sector of the diagram, all but one with OI-type spectral profiles, i.e.\ \DPS\ profiles with negative radial gradients, which could be the result of combining a weak central AGN with little or no extended star formation. Meanwhile, the distribution of the LINER S0s runs throughout the three activity regimes, but showing a tendency to concentrate in the vicinity of the PS with their peak straddling the PS--TR divide. In regard to the star formation, these are galaxies located mainly in the GV, although there are also a fair amount of quiescent objects and some in the MS as well. 

\begin{figure}
    \centering
	\includegraphics[width=\columnwidth]{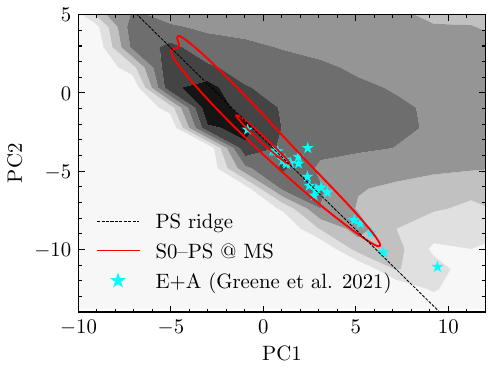}
    \caption{Location of the SDSS spectra of the sample of E+A galaxies (cyan stars) studied by \citet{Greene+2021} in the PC1--PC2 plane defined in \citealt{Tous+2020} for present-day S0 galaxies. The red contours enclose $10$ per cent (inner) and $90$ per cent (outer) of the probability of finding MaNGA PS--S0 systems included in the MS of SF galaxies. The dashed line shows the ridge of the PS given by equation~(\ref{eq:ps}). The greyscale contours in the background are the same as Fig.~\ref{fig:ps_pc1_pc2}.}
    \label{fig:ea}
\end{figure}

Likewise, although most of the `blank' S0s are, as it could not be otherwise, passive, quiescent galaxies, it can be observed that some of them reach the GV, and a few even the MS. Within this last subset, we find lenticulars with relatively high SSFRs ($\gtrsim 10^{-11}\; {\rm yr^{-1}}$), as well as distributions of stellar mass and stellar mass-to-light ratio in the $r$-band centred at slightly smaller values than the bulk of the PS population. From the spectral point of view, these are passive objects whose optical spectrum has a fairly blue continuum, lacks obvious emission lines and shows strong Balmer absorption lines. These characteristics are found in the galaxies that populate the lower half of the PS ridge (see, for instance, panels d and e of fig.~B1 in \citetalias{Tous+2020}). We suspect that these PS--S0s located in the MS are probably mostly post-starburst E+A galaxies \citep{Dressler+1992}. This is, in fact, what seems to emerge from the PC1--PC2 diagram of Fig.~\ref{fig:ea}, where we draw the location of the single-fibre SDSS spectra of the sample of E+A galaxies studied by \citet{Greene+2021}. Remarkably, almost one third of this E+A sample lies within the $90$ per cent contour of the peak density of the distribution of the MaNGA PS--S0s that are classified in the MS (inner red curve), while only three of them fall outside the $10$ per cent boundary of this distribution (outer red curve). The presence of E+A galaxies within the SF MS might initially appear paradoxical, as their lack of emission line indicators suggests little-to-no ongoing star formation. However, this apparent contradiction can be reconciled by considering that the values of the SSFR are inferred over a specific timescale. In particular, the SSFRs derived in this work from \citeapos{Salim+2018} data account for the star formation activity over the past $100$ Myr.

% Fig. 5

\begin{figure}
    \centering
	\includegraphics[width=\columnwidth]{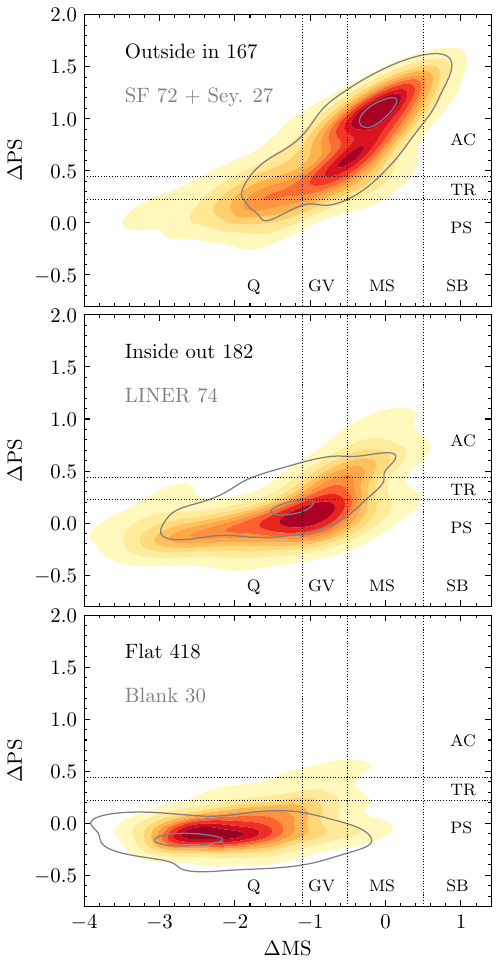}
        \caption{Kernel density estimates of the distributions in the \DPS--\DMS\ diagram of the MaNGA S0 sample divided into three groups according to the sign of the radial gradient of activity, from top to bottom panels: OI, IO and F. The intensity of the coloured contours positively correlates with the number density of data points. The grey contours on top of the distributions enclose $10$ per cent (inner) and $90$ per cent (outer) of the probability of finding MaNGA S0s classified by \citet{Thomas+2013} as SF or Seyfert (top panel), LINER (middle panel), or that could not be classified because their emission lines were unavailable (bottom panel). Subset sizes are indicated next to each label. The horizontal and vertical dotted lines show the same demarcations as Fig.~\ref{fig:bpt_dps_dms}}
        \label{fig:dps_dms_quenching}
\end{figure}

We complete this subsection by showing in Fig.~\ref{fig:dps_dms_quenching} the distributions of the galaxies of the MaNGA S0 sample in the \DMS--\DPS\ plane according to the sign of their activity gradient. This plot shows, on one hand, that there is a clear link between the peak density of the three distributions and the star formation status of the galaxies. Thus, it can be seen that while OI-type profiles occur in S0s with highly diverse but positively correlated levels of activity and star formation, the bulk of these objects populates the MS--AC sector of the subspace. Likewise, among lenticular galaxies with IO activity profiles, PS class objects abound, which can nevertheless show in some cases levels of star formation typical of the MS, so the peak of this distribution is located at the top of the GV--PS sector. For their part, virtually all galaxies with F gradients are Q--PS systems that have stopped forming stars and show no other signs of activity. Note that this relationship between the star formation status of S0s and the sign of their radial gradient of activity is a more than remarkable result, because the former is inferred via equation~\ref{eq:DMS} from two observables, SFR and $M_\ast$, retrieved from the GSWLC-2 catalogue, where they were determined by multi-SED fitting, while the shapes of the \DPS\ profiles come from a completely independent PCA carried out on the MaNGA data cubes. On the other hand, Fig.~\ref{fig:dps_dms_quenching} also reveals that the distributions of the activity gradients substantially overlap with the distributions corresponding to the activity classes inferred from the conventional \ion{N}{II}--BPT diagnostic. Since the first of these distributions present a high degree of mutual overlap in the \DMS--\DPS\ plane, the existence of a common ground in said subspace between BPT types and the sign of activity gradients does not guarantee per se that both classifications are closely related. Even so, we find that 78 per cent of the SF and Seyfert S0s have OI profiles -- actually all but one of the Seyfert show this type of profile --, while 73 per cent of those that do not fit in any of the BPT categories according to \citet{Thomas+2013} due to the lack of emission lines have flat \DPS\ profiles. Only in the case of LINER S0s do we obtain a somewhat more balanced distribution of activity morphologies: 49 per cent F, 30 per cent IO, and 22 per cent OI. However, a good number of them (49) are included in our spectral class PS. Inspection of their $M_\ast$ and EW(\Ha) distributions shows that virtually all (48) have $M_\ast/M_\odot\geq 10^{10}$ and EW(\Ha)\;$\leq 3$. And what is most interesting, all but 3 LINER S0s with an optical spectrum of the PS class possess either IO (15) or F (31) radial activity profiles. Such features lead us to argue, in agreement with the findings in \citet{JimP+22}, that any trace of emission displayed by these systems will most likely be powered not by nuclear accretion, but by photoionization by very hot post-AGB stars of any diffuse gas they withheld \citep[e.g.][]{Cid+2010}. Again, these are results that can be described as remarkable, since the radial gradient of \DPS\ does not play any role in the BPT classification (or lack thereof), which is based on the integrated optical spectra of galaxies. 

% Fig. 6

\section{Activity profiles vs physical properties}
\label{S:profiles_vs_properties}
Here, we offer an overview of the relationship of the radial distribution of the activity in the different spectral types of present-day S0s with several physical properties that act as hallmarks of galaxy formation mechanisms.

\subsection{Total stellar mass and local density of the environment}
\label{SS:dps_mstar_density}
\begin{figure*}
    \centering
    \begin{subfigure}{0.88\textwidth}
        \includegraphics[width=\linewidth]{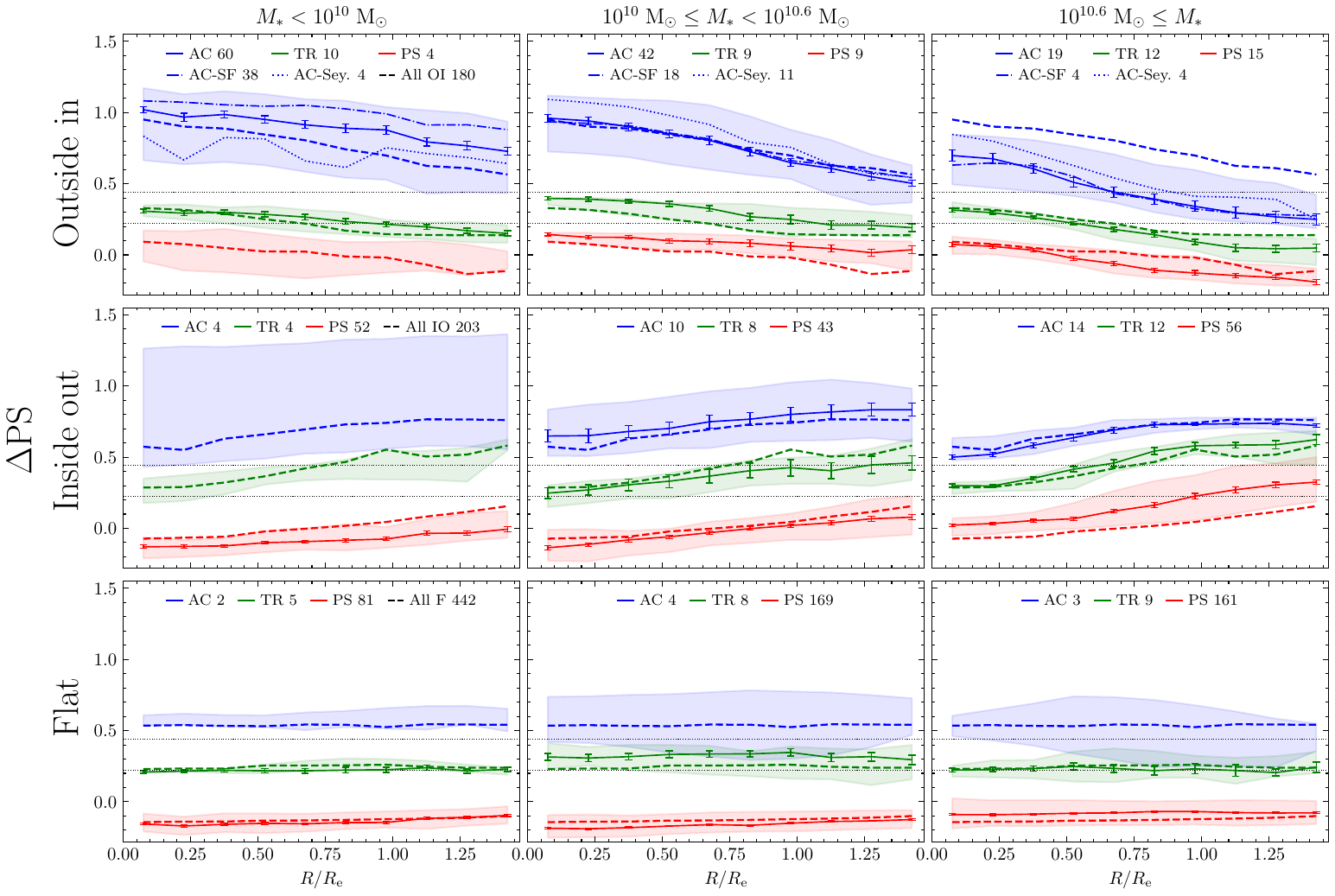}
    \end{subfigure}\\
    \begin{subfigure}{0.88\textwidth}
        \centering
        \includegraphics[width=\linewidth]{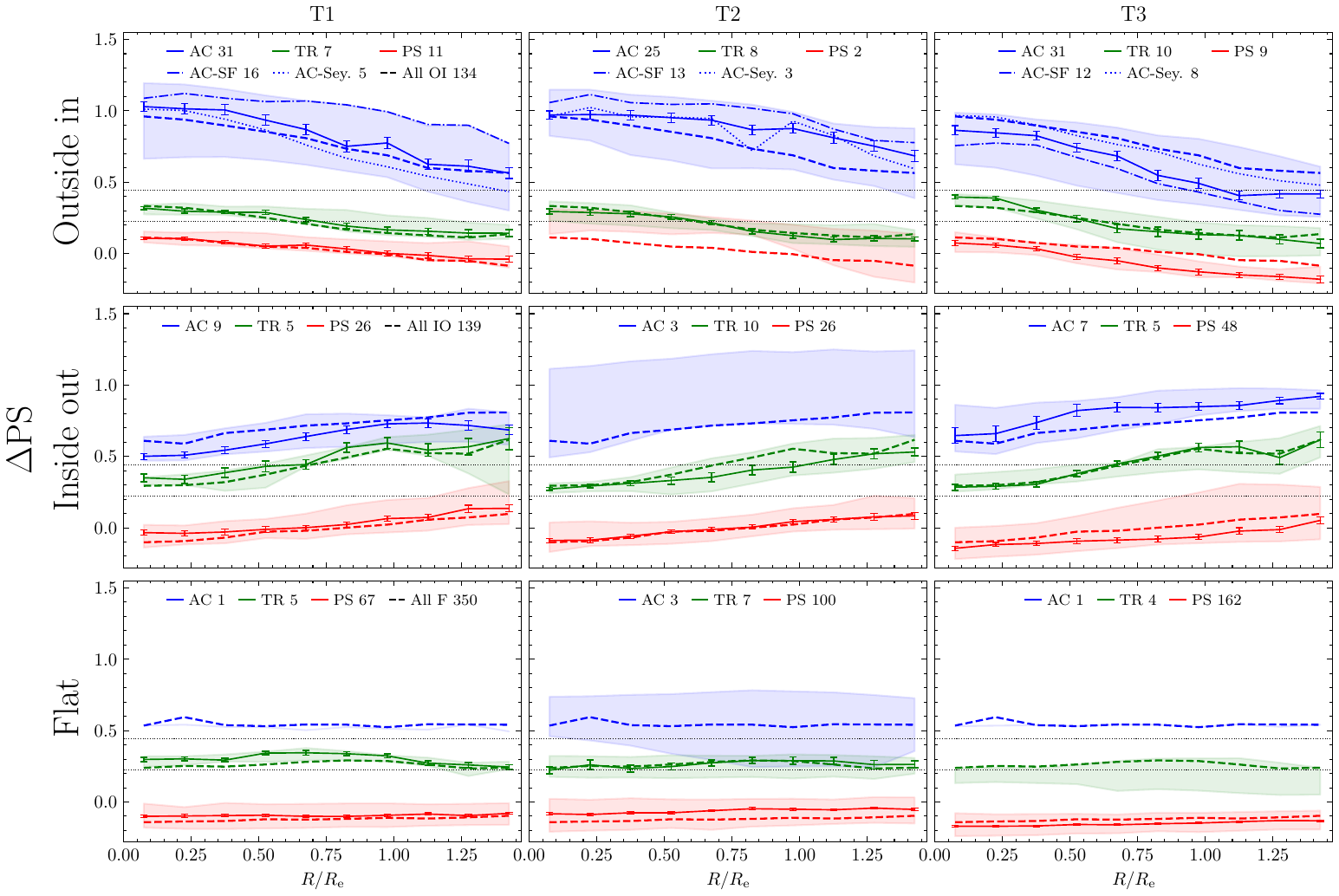}
    \end{subfigure}    
    \caption{\emph{First three rows of panels:} median radial \DPS\ profiles and $1\sigma$ uncertainty (solid curves and error bars) of present-day AC (blue), TR (green) and PS (red) S0 galaxies divided into three bins of $M_\ast$ and by the sign of their radial gradient of activity: OI (top), IO (middle), and F (bottom). Shaded regions show the interquartile range of the profile distribution for samples with more than four objects; otherwise, they show the total area encompassed by all the available profiles. For the AC:OI subset, we also show the spectral profiles of galaxies classified by \citet{Thomas+2013} as star forming (SF, blue dash-dotted line) and Seyfert (Sey., blue dotted line). The median radial profile of all galaxies of each spectral class with the same activity gradient irrespective of $M_\ast$ (dashed curves) is repeated along the panels of the same row for reference. The horizontal dotted lines are the boundaries of the TR zone. The numbers of objects used in deriving the statistics of each subset are indicated in the legends. \emph{Last three rows of panels:} as upper rows, but grouping the radial profiles into terciles of local density, T1, T2 and T3, which roughly correspond to low-, medium- and high-density environments.}
    \label{fig:dps_profiles_prop}
\end{figure*}

The stellar mass is quite possibly the key intrinsic property in the evolution of galaxies. To assess its relationship with the radial run of activity in S0 galaxies, we have represented in the top nine panels of Fig.~\ref{fig:dps_profiles_prop} the median \DPS\ profiles of the three spectral types of these objects (i.e.\ AC/TR/PS) arranged in three rows according to the sign of the gradient and in three columns corresponding to $M_\ast$ grouped in disjoint bins. As this figure shows, the largest subgroup within the S0s of the AC class (blue solid curves and/or associated shaded regions) is that of low-$M_\ast$ objects with OI quenched activity profiles, while their PS counterparts (red solid curves and/or associated shaded regions) prefer higher masses and F activity profiles. Conversely, TR lenticulars (green solid curves and/or associated shaded regions) are relatively evenly distributed, both in terms of $M_\ast$ and the sign of the activity gradient. Using as reference the median radial profiles of all galaxies with the same gradient and class of activity calculated ignoring the division in mass (dashed curves), we can also observe in Fig.~\ref{fig:dps_profiles_prop} that the characteristic level of activity of the AC galaxies with OI profiles decreases at all galactocentric radii as $M_\ast$ increases. This bearing is produced by the progressive reduction of the SSFR with mass in the SF S0s included within this class (see the blue dot-dashed curves in the panels of the top row), a result that is in good agreement with the findings of e.g.\ \citet{Belfiore+2018} for the general disc-galaxy population. For their part, the Seyfert-type S0s show a more erratic relationship with $M_\ast$ (blue dotted curves, but note the small number of these objects) which makes their entire median activity profile to run clearly below that of the SF in the lowest-mass bin ($M_\ast/M_\odot < 10^{10}$) and, instead, reverse this trend at intermediate and high masses. In the same way, one can also see that the negative gradients of the median \DPS\ profiles of the AC:OI--S0s also tend to become steeper with mass. This effect, combined with the gradual reduction in activity intensity, causes much of this profile to penetrate deeply into the TR in the highest mass range. Despite the smallness of the subsamples, the above trends can also be guessed in the characteristic profiles of PS--S0s with OI activity gradients (red curves in the upper row of panels of Fig.~\ref{fig:dps_profiles_prop}). 

Focussing on the median \DPS\ profiles of type IO included in the second row of panels of Fig.~\ref{fig:dps_profiles_prop}, those associated with PS systems also show a progressive steepening of the -- in this case positive -- gradient with $M_\ast$, their level of activity increases instead of decreasing. This leads the outermost section of this profile to cross, in the high-mass bin, the upper boundary of the PS class and enter the TR. Regarding the AC:IO profiles, the most remarkable feature is not so much related to their shape but to the fact that their fractional abundance within this spectral class, contrary to what happens with those with OI gradients, increases with $M_\ast$. A similar result, but restricted to the quenching of star formation, has been reported by \citet{Lin+2019} for the general population of emission-line galaxies.

Probably due to their intermediate spectral classification, the different types of TR profiles exhibit a somewhat less defined trends with mass, although with a certain preference to imitate those from their PS counterparts. As shown by the green profiles in the upper subset of nine panels of Fig.~\ref{fig:dps_profiles_prop}, the outer parts of these profiles usually go into one of the adjacent spectral regions, AC or PS depending on the positive or negative sign of the gradient, respectively, while the strength of the activity does not seem to follow a defined pattern. As for the S0s with F-type profiles, represented in the third row of panels, no significant variations with $M_\ast$ are observed in any of the spectral classes, the most notorious result being that, independently of the value of this variable, $\gtrsim 75$ per cent of the PS:F--S0s have profiles with \DPS\;$<0$ in the entire explored range of galactocentric distances. 

A similar procedure is applied to determine the part played by the density of the environment, the most important external factor when talking about evolution, in the activity profiles of S0 galaxies. To this effect, we show in the last three rows of panels in  Fig.~\ref{fig:dps_profiles_prop} the distributions of the \DPS\ profiles of each activity class divided into three blocks, or terciles, according to the rank reached by the best estimate of the local density for each galaxy: columns T1, T2 and T3, from lowest to highest. Here too, galaxies with the same type of activity gradient are depicted in the same row. It can be seen, that the median profiles of the AC and TR activity classes exhibit behaviours with increasing local density that, to some extent, are not too different from those they showed with increasing mass. If anything, the reduction in the strength of activity of the AC:OI profiles (first row of the lower subset of nine panels in Fig.~\ref{fig:dps_profiles_prop}) is now less progressive and pronounced. This is so to the point that the median \DPS\ profile in the third density tercile barely penetrates the TR in the densest environments, in this case probably due to the greater relative contribution to the activity level of the entire AC subset, both in number and magnitude, of the Seyfert S0s against the SF S0s. On the other hand, the median TR--S0 profiles with a non-null gradient are characterized by penetrating at high $R$, and regardless of environmental density, into the adjacent spectral zones, reproducing the same behaviour observed with $M_\ast$. Meanwhile, the typical profiles of the PS:IO--S0s (second row of the lower subset of panels) do not show significant variations with the environment, remaining always entirely within their activity class, although for the few AC--S0s with IO gradients the average level of activity does seem to increase slightly with density. This would come to confirm that, unlike the $M_\ast$, the density of neighbouring galaxies does not play a relevant role in the existence of IO profiles in S0s, which are usually associated with SF rings \citetext{\citealt{Tous+2023}; see also Section~\ref{SSS:mini_mergers}}. We also observe in the bottom row of Fig.~\ref{fig:dps_profiles_prop} that the F-type profiles classified according to this extrinsic variable, in complete agreement with the results derived from the sampling based on $M_\ast$, are those that display a more neutral behaviour for all spectral classes. And there is also coincidence in the fact that in all density ranges the entire profiles of around three out of four PS:F systems exhibit negative \DPS\ values.

%Fig. 7

\begin{figure}
    \centering
    \includegraphics[width=\columnwidth]{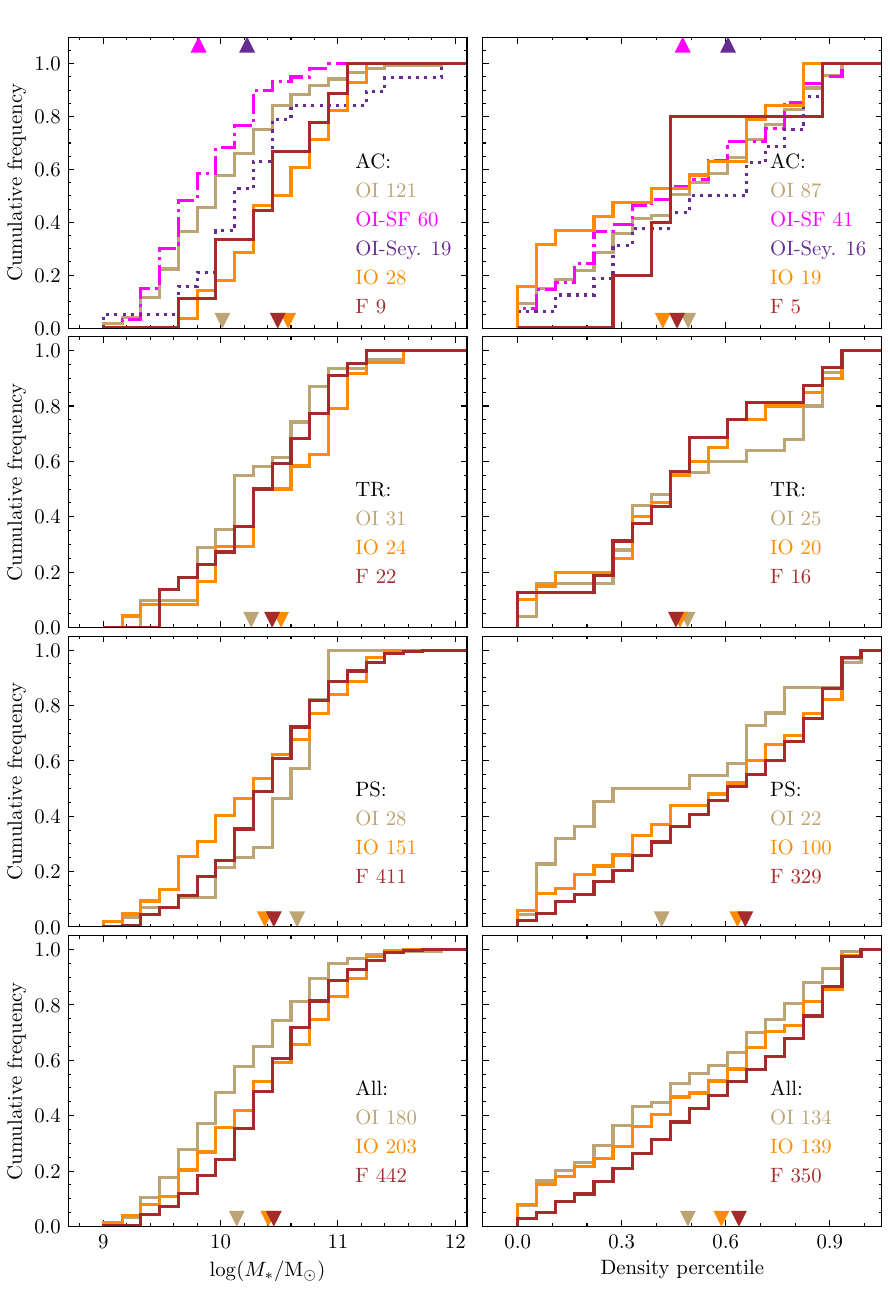}
    \caption{\emph{Left panels:} cumulative distribution functions of $M_\ast$ for the different types of activity profiles of the MaNGA S0 galaxies. The solid lines show outside in (OI, light brown), inside out (IO, orange) and flat (F, brown) gradients. Each panel depicts the CDFs segregated by activity class, from top to bottom: active cloud (AC), further divided into galaxies with a SF (dash-dotted line) and Seyfert (dotted line) status assigned by \citet{Thomas+2013}, transition region (TR), and passive sequence (PS), respectively, while the bottom panel shows aggregate results. The filled triangles indicate the medians of each CDF. The size of each subset is shown in the legend next to the profile labels. \emph{Right panels:} as the panels on the left, but using as random variable the percentile corresponding to the local density of the environment (see Section~\ref{S:data}).}
    \label{fig:cdf_mstar_density}
\end{figure}

The information conveyed by Fig.~\ref{fig:dps_profiles_prop} is complemented by Fig.~\ref{fig:cdf_mstar_density}, which shows the empirical cumulative distribution functions (CDFs) of $M_\ast$ (left column of panels) and environmental density (right column of panels), corresponding to the subsamples of each type of activity gradient, both divided into activity classes and aggregated. The upper-left panel reveals that within the lenticulars of the AC class there is a large difference between the CDFs of galaxies with OI and both IO and F gradients, the latter two being notably more massive in statistical terms. Comparison of the mass distribution of the subsample of OI gradients with the other two via a two-sample Kolmogorov-Smirnov (KS) test shows, however, a statistically significant difference only with respect to the IO subsample, for which the test returns a low $p$-value ($<< 0.05$), indicating that it is very unlikely that the masses of the AC:IO--S0s arise from the same parent distribution than those of the corresponding OI subset. For the AC:F subset, its small size (9 galaxies) prevents the KS test from giving a significant result. Nonetheless, we go on to obtain highly significant differences in all cases when  these comparisons involve those AC:OI with an SF BPT-classification, something that also occurs when the Seyfert and SF subsets are compared to each other. This allows us to objectify the fact that the latter have the smallest masses of all S0s. 

For the lenticulars of the PS class significant differences are also found between the CDF of the IO profiles and those of the other two types of gradient (third panel), although in this case the order of the median values of the mass is reversed, so the PS:IO--S0s are now somewhat less massive than the rest on average. In contrast, and also likely due to the relative smallness of the different subsamples, we find no statistically meaningful differences among the CDFs of the three types of gradients within the TR population (second panel). On the other hand, the larger size of the samples of the aggregated distributions that are obtained by not partitioning the data into activity classes and that we have depicted in the bottom-left panel of Fig.~\ref{fig:cdf_mstar_density}, makes them all significantly different from each other according to the two-sample KS test. The order of the three subsets ranked by the average $M_\ast$ of their member galaxies is OI, IO and F, from lower to higher.

%Fig. 8

On the other hand, the shapes of the CDFs of the local density percentile shown in the upper-right panel of Fig.~\ref{fig:cdf_mstar_density} hint at the possibility that the AC:IO--S0s are somewhat more abundant in low-density galaxy environments than their OI counterparts and, especially, than the F ones, which almost completely avoid the first density tercile (but beware of the small number of galaxies (5) of this subclass). Nevertheless, and in accordance with the similarity shown by the median values, none of the observed differences turns out to be statistically significant, not even if we involve the AC:OI SF and Seyfert in the comparisons -- even though the latter show a clear predilection for residing in denser environments. In the case of the PS--S0s (third panel) it is the CDF of the OI profiles that is positively skewed with respect to the other two distributions, although according to the two-sample KS test, only the comparison between the OI and F subsamples of these passive systems yields a statistically significant difference. And just as it happens with the mass, we cannot reject either for the TR galaxies (second panel) the null hypothesis that the three subsamples of gradients come from the same parent distribution of local galaxy densities. The similarities between the CDFs of $M_\ast$ and environmental density also extend to the ascending ordering of the median values of the samples of the aggregate profiles, which for this extrinsic variable is again OI, IO and F. In this case, however, the differences among the CDFs of the aggregated profiles (bottom-right panel of Fig.~\ref{fig:cdf_mstar_density}) turn out to be  statistically significant only between the OI and F subsets.
    
\subsection{Stellar ages and kinematics}
\label{SS:dps_age_kinematics}
We now turn to investigating if there is any connection between the radial gradients of activity of S0 galaxies and the ages and kinematics of their stellar populations. 

\begin{figure}
    \centering
    \includegraphics[width=\columnwidth]{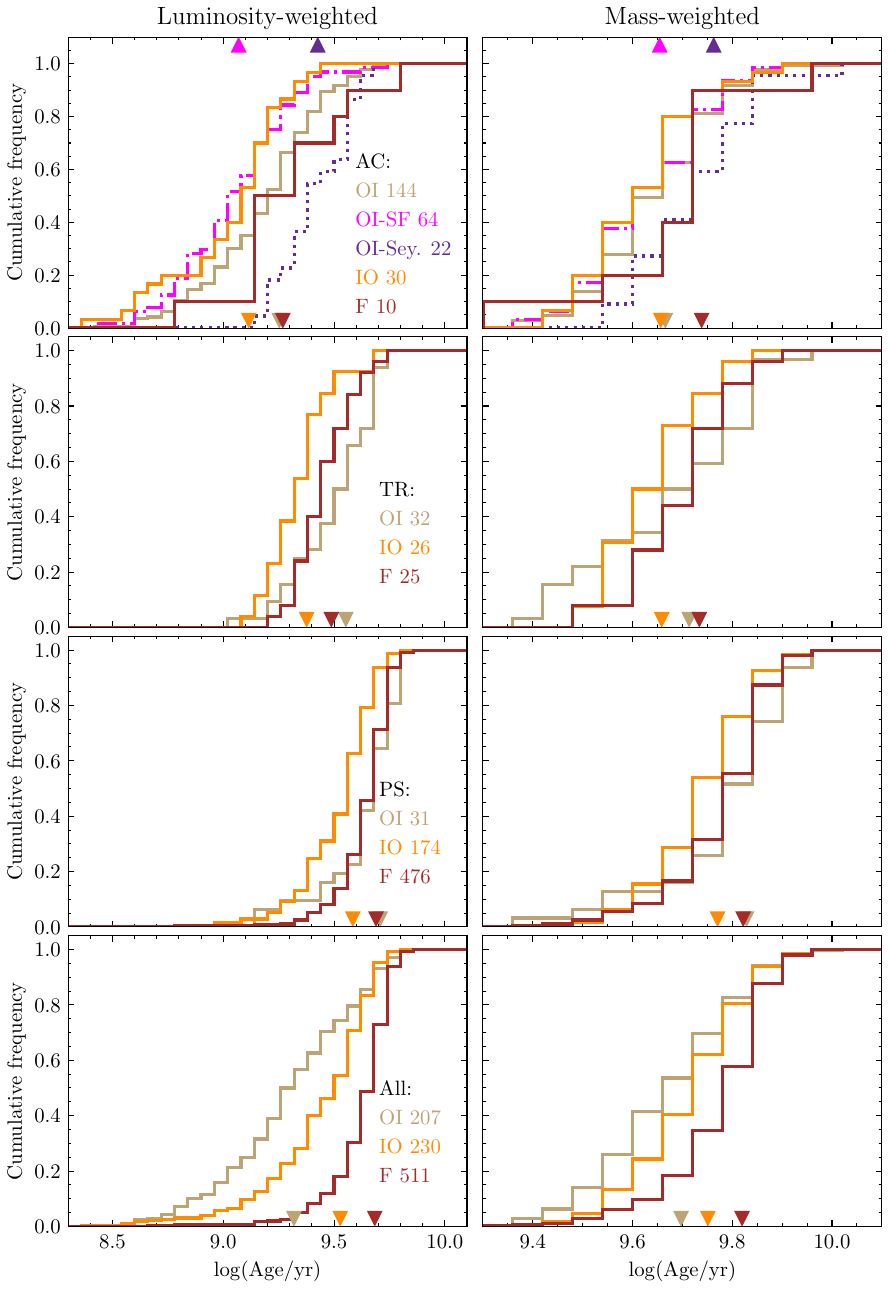}
    \caption{As Fig.~\ref{fig:cdf_mstar_density} but using as random variables the luminosity-weighted age (left panels) and mass-weighted age (right panels) of the stellar populations.}
    \label{fig:cdf_age}
\end{figure}

To delve into the dependence with stellar age, we apply the same methodology used in Fig.~\ref{fig:cdf_mstar_density}, so we compare in Fig.~\ref{fig:cdf_age} the CDFs of the mean stellar age of S0s weighted with both luminosity (left column of panels) and mass (right column of panels) for the different activity gradients. As before, the upper three panels collect from top to bottom the results corresponding to the AC, TR and PS classes of activity, respectively, while the lower ones show aggregated results. In the case of the luminosity-weighted ages, the spectral class AC is, as would be expected, the one that harbours, regardless of the sign of the gradient, the youngest stellar population on average, followed by the TR systems and then the PS systems. In this panel, the most discrepant and significantly different CDF corresponds to the Seyfert AC:OI subgroup, which contains the galaxies with the oldest luminosity-weighted stellar populations within this spectral class. This result agrees perfectly with the idea, already expressed in Section~\ref{S:activity_profiles}, that the presence of an AGN has a negative impact on the SFRs of the S0s. Fig.~\ref{fig:cdf_age} also shows that, in each of the three spectral classes, lenticular galaxies with IO activity gradients harbour fractions of younger stars -- which are the ones that dominate the optical spectrum -- somewhat larger than those of the other two types of profiles, which translates into significant discrepancies with the CDFs of both when applying the two-sample KS test. This finding implies that the relative contribution of recent star formation to overall activity, regardless of its level, is more important in S0 galaxies with IO profiles than in their OI and F counterparts. Only in the bottom panel depicting the cumulative histograms of the integrated light-weighted mean ages, all three gradients show distinctly segregated CDFs which in this case are ranked OI, IO and F from lowest to highest ages. This apparent development in the sequence of ages of the stellar populations simply takes account of the fact that the bulk of S0s with OI profiles are SF galaxies of the AC class, while most systems with both IO and F activity gradients belong to the quiescent PS class, just as shown in Fig.~\ref{fig:dps_dms_quenching}.

The CDFs of the mass-weighted mean stellar ages of the MaNGA S0s are depicted in the right column of panels of Fig.~\ref{fig:cdf_age}. In addition to an overall increase in the ages, explained by the fact that mass-weighted values are a good proxy for the integrated star formation history of galaxies and, therefore, of the actual average age of their stellar populations, we can now verify that the distributions are, in general, much more similar to each other within a given spectral class (note the different scale of the abscissae with respect to the left column). The only exceptions are the Seyfert AC:OIs, which again host stellar populations significantly older than their SF AC, IO and F counterparts, and the PS:IOs, for which the $p$-values from the KS test suggest that their mass-weighted stellar ages could come from a slightly younger parent population than that of the other S0s belonging to this passive subset. The broad homogeneity shown by the characteristic stellar ages of S0s regardless of their level of activity gives way in the last panel to small, but statistically significant differences in the aggregated CDFs, which reproduce the same arrangement of median ages $\mbox{OI} < \mbox{IO} < \mbox{F}$ observed in the corresponding light-weighted measurements. The correlation of the sign of the activity gradients with the SSFRs (Fig.~\ref{fig:dps_dms_quenching}) tells us that this sequence is due to the fact that we are dealing with old galaxies that harbour different fractions of \emph{newly formed} stars. Indeed, the results of this particular analysis provide a strong argument in favour of considering active S0s to arise primarily from the rejuvenation of their passive counterparts. However, as will be discussed later (Section~\ref{S:discussion}), some of our findings also support, or do not contravene, the fading of Sps as a viable explanation for the existence of AC--S0s.

%Fig. 9

% Fig. 10
\begin{figure*}
    \centering
    \includegraphics[width=0.95\textwidth]{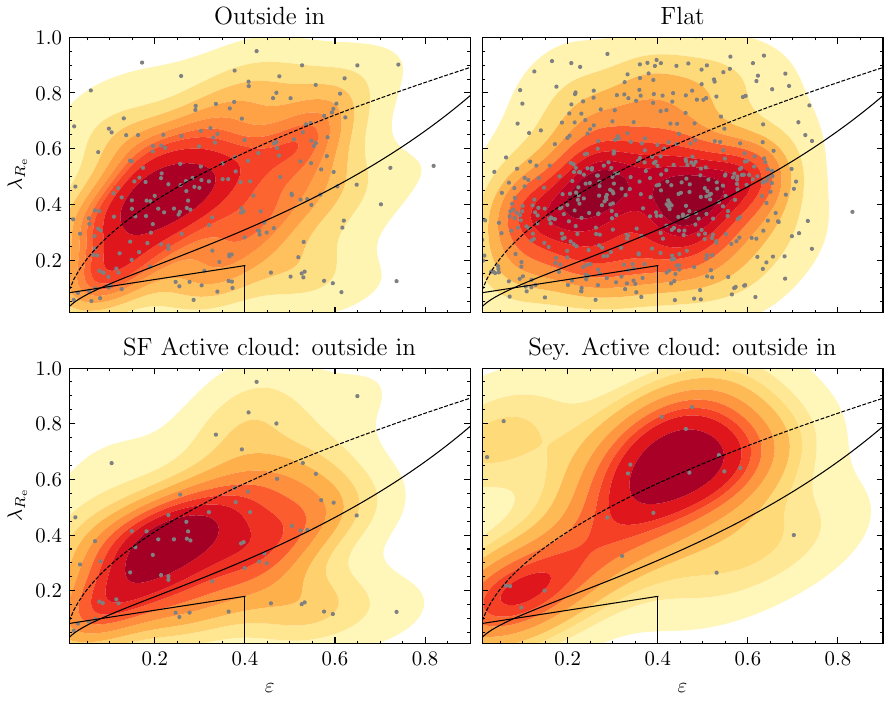}
    \vspace{-0.3cm}
    \caption{Kernel density estimates of the distributions of the MaNGA S0 galaxies with OI or F activity gradients in the $(\lRe,\varepsilon)$ diagram (top panels), and of those members of the AC:OI subset classified as SF or Seyfert by \citet{Thomas+2013} (bottom panels). The darker the colour, the higher the number density of data points. The grey dots show the individual values. The luminosity-weighted averaged projected specific angular momentum of the galaxies, $\lRe$, is calculated from equation~(\ref{eq:angular_momentum}). The black curves, the same in all panels, indicate the theoretical relations expected for oblate edge-on models of regular rotators that are either isotropic, $\beta = 0$ (top dashed lines), or have an anisotropy $\beta = 0.7 \varepsilon_{\rm intr}$ (bottom solid lines). The small boxes in the lower-left corner provide the best empirical fast/slow rotator divide, as given by equation~(19) in \citet{Capellari2016}.}
    \label{fig:lambda_epsilon}
\end{figure*}

To investigate possible dependences of the radial gradients of activity with the stellar kinematics of S0s, we resort to examining the bivariate distributions of the central values of the stellar spin parameter, $\lRe$, defined in Section~\ref{S:intro}, and the overall observed ellipticity, $\varepsilon$, for the various subsets of these galaxies that we have defined along this work. Fig.~\ref{fig:lambda_epsilon} compares the kernel density estimates of four of these distributions in the $(\lRe,\varepsilon)$ diagram: the S0s with OI or F activity gradients and those identified as SF or Seyfert by \citet{Thomas+2013} within the AC:OI subclass\footnote{To make this plot more compact, the diagram corresponding to the IO activity gradients, which depicts an intermediate behaviour between the OI and F cases, has not been included. Nor are the diagrams of the AC, TR and PS spectral classes shown, since they do not provide relevant new information. All of them are, however, discussed in the text.}. To guide the eye, we have included in all four panels curves showing the theoretical relationship expected from the dynamical modelling of isotropic regular rotators (top dashed line), as well as from those obeying the lineal relation between anisotropy and ellipticity given by \citep[cf][]{Capellari+2007} $\beta = 0.7 \varepsilon_{\rm intr}$ (bottom solid line), in both cases for galaxies seen edge-on\footnote{A rigorous application of this anisotropy diagram and the different guide lines adopted assumes that the summation~(\ref{eq:angular_momentum}) used to calculate $\lRe$ is not spatially limited and that the value of $\varepsilon$ is intrinsic and constant with radius. However, the fact that we are only interested in an internal comparison between different S0 subsets, whose unknown selection functions are expected to be uncorrelated with projection effects, allows us to carry it out directly using the observational estimates of both parameters.}. Randomly oriented real fast rotators are expected to broadly form an envelope to the left and above this latter line, which acts as an empirical boundary for anisotropic rotators structurally flattened by rotation, while counter-rotating disc galaxies will tend to fall below it. Besides, we have added in the lower-left corner of each individual diagram a box providing the best empirical separation between fast and disc-less slow rotators, as given by equation~(19) in \citet{Capellari2016}, and that encloses mainly, but not all, E galaxies. 

Despite the large scatter usually associated with individual measurements, diagrams based on stellar kinematics do a rather good job separating the two main groups of fast and slow rotators when they are applied to the general galaxy population \citep[e.g.][]{Emsellem+2007}. However, the sensitivity of these plots becomes rather limited when it comes to detecting differences among members of the family of fast rotators, especially if they are of the same Hubble type, as in the present case. Nevertheless, as depicted in the top panels of Fig.~\ref{fig:lambda_epsilon}, the $(\lRe,\varepsilon)$ diagrams of the two aggregate distributions of the OI and F gradients, reveal some eye-catching differences that are also broadly replicated by the distributions of their main spectral contributors, the activity classes AC and PS, respectively. Thus, while both distributions cover quite similar ranges of the parameter space and tend to avoid the region associated with slow rotators (see also Appendix~\ref{A:morphological_contamination}), their shapes are clearly distinct. In the case of the OI subset (left panel), much like the AC, the data shows a single peak centred on top of the upper guide line, in the region of the diagram bounded by the coordinates $0.15\lesssim\varepsilon\lesssim 0.35$ and $0.3\lesssim\lRe\lesssim 0.5$, accompanied by traces of a possible secondary peak at higher values of both $\varepsilon$ and $\lRe$ (see below and Appendix~\ref{A:morphological_contamination}). On the other hand, the F subset (right panel), as well as the PS, showcase a clearly bimodal distribution, with the maxima relatively close to each other and located between the two divide lines in a region where $\varepsilon$ ranges from $\sim 0.2$ to $0.55$ and where the modal values of $\lRe$, as happens with the OI and AC subsets, remain centred around $0.4$, which is typical of nearby lenticular systems \citep{Capellari2016}. For their part, the IO and TR subsets, in which GV galaxies abound, exhibit essentially unimodal distributions with maxima located approximately between those of the extreme cases depicted in the plot. Objective comparison of all these distributions using a 2D KS test\footnote{The \textsc{python} implementation of this test is available at \url{https://github.com/syrte/ndtest} (not to be confused with the two-sample KS test used above to compare pairs of CDFs).}, shows that the differences between the OI (AC) and F (PS) subsets are in fact significant. 

To investigate the origin of the above  discrepancies, we have also inferred the distributions of anisotropies of the two main contributors to the AC:OI subgroup of S0s. As can be seen in the lower-left panel of Fig.~\ref{fig:lambda_epsilon}, the distribution of the AC:OI lenticulars classified as SF only presents minor differences with respect to that of all OI systems, which are specified in a slight displacement of the maximum towards lower values of $\lRe$ and the total absence of evidence of a second peak, something that is not unexpected given that these objects are the most abundant in this type of gradient. By contrast, the distribution of the Seyfert S0s with AC:OI profiles in the lower-right panel shows a small peak near the origin and a primary maximum notably shifted towards higher values, not only of $\varepsilon$, which now overlap with those of the second maximum of PS:F systems, but also of $\lRe$, that exceed those typical of the S0 population. This suggests that within S0s, AGN episodes take place preferentially in galaxies that have more rotational support than most. Although the small size of our Seyfert's subset (21 objects) limits the robustness of this latter result, we note that \citet{Moral-Castro+2020} have recently reached a similar conclusion based on an analysis of the dimensionless stellar spin parameter too, but applied to isolated Sa--Sbc type discs (see also Section~\ref{SS:hydrodynamics}).

\section{Discussion}
\label{S:discussion}
In this section, we aim to assess the alignment between the predictions of the main evolutionary channels for local S0s and the results of our analysis on their activity profiles, particularly highlighting the role played by hydrodynamic interactions and gravitational encounters in the formation of these objects.

\subsection{Activity driven by hydrodynamics}
\label{SS:hydrodynamics}
Hydrodynamic mechanisms, exemplified by ram pressure stripping \citep[RPS,][]{Gunn&Gott1972} and accompanying transport processes like turbulent viscous stripping \citep{Nulsen1982,RH2005}, entail high-speed interactions between the galaxies' ISM and a dense and hotter IGM, two conditions that are met in large aggregations of these objects. Candidate S0 galaxies that may have acquired their morphology by this mean are former spiral galaxies that transitioned after falling into a cluster or a rich group, preferably during the making of these systems \citep{Solanes+2016}.

IGM-ISM interactions in galaxies primarily result in the removal, occurring on time-cales of $\sim 1$--$2\times 10^9\;$yr \citep{Lotz+2019}, of both the cold atomic gas in the discs and the surrounding hotter gas reservoir (strangulation), leading to an overall outside-in reduction in star formation activity and a potential localized increase in the SFR at the ISM-IGM interface, with the complete transformation into an S0 necessitating of a subsequent galaxy-wide structural rearrangement induced either externally or by the sweeping gas itself \citep[][]{Lee+2022}. While the gravitational field of massive galaxies may be strong enough to retain some neutral gas in their central regions or to reaccrete a certain fraction of the displaced ISM, in small galaxies the stripping of the \ion{H}{I} may be total and irreversible, which may accelerate their transition to quiescence \citetext{see \citealt{Boselli+2022} and references therein}. Conversely, the impact of RPS on the nuclear activity of galaxies remains unclear. Some authors regard clusters as a hostile environment for AGNs \citep[e.g.][]{Dressler+1985,Lopes+2017,Mishra+2020}, whilst others argue that hydrodynamic interactions could actually enhance the feeding of the central SMBHs \citep[e.g.][]{Ricarte+2020,Peluso+2022}. 

\begin{figure*}
    \centering
	\includegraphics[width=0.95\textwidth]{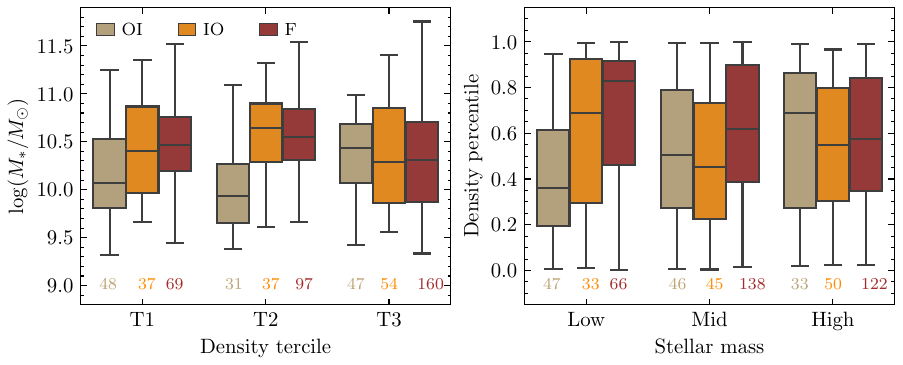} 
    \vspace{-0.3cm}
    \caption{Box plots summarizing the aggregate distributions of stellar mass for the OI, IO and F types of the radial gradient of activity in each of the three terciles of environmental density (\emph{left panel}) and, vice versa, of the density percentile of such gradients in the three bins of stellar mass defined in Fig.~\ref{fig:dps_profiles_prop} (\emph{right panel}). The figures underneath each box show the number of objects included in the different subsamples.}
    \label{fig:box_percentile_mstar}
\end{figure*}

This scenario certainly calls for S0s with OI profiles (and possibly F profiles too). As shown by Fig.~\ref{fig:dps_profiles_prop}, the level of activity of galaxies with this type of gradient diminishes with increasing stellar mass and/or environmental density, leading to significantly reduced \DPS\ values in the outer radial bins. This trend is accompanied by a rise in the relative abundance of PS (and TR) OI lenticulars at the expense of systems featuring an AC:OI configuration, which generally tend to be less massive and younger, especially those classified as SF (Figs.~\ref{fig:cdf_mstar_density} and \ref{fig:cdf_age}). The substantial difference in typical masses is precisely what prevents the former from being broadly considered as the result of the fading of the latter. Figure~\ref{fig:box_percentile_mstar} complements previous plots by contrasting summarized stellar mass distributions across three local density terciles (left panel) and vice versa, density percentile distributions across three stellar mass bins (right panel), with spectra categorized solely by gradient type to ensure a reliable statistical analysis. While it is obvious that in the second panel the differences between the distributions are less pronounced, they still allow us to discern certain trends, such as, for instance, that the central part of the distribution of local densities for galaxies exhibiting OI gradients (depicted by light brown boxes) undergoes a slight shift toward higher values as mass increases. This indicates a tendency for the most massive S0s with this activity profile to be more commonly found in densely populated regions.

%Fig. 11

Despite the lack of significant variations in the \DPS\ profiles of the Seyfert AC--S0s with the local density, our finding that Seyfert S0s tend to be more massive, elongated and rotationally supported than their SF counterparts (see Fig.~\ref{fig:lambda_epsilon}), can still be seen as evidence favourable to the contribution of RPS to AGN activity. A similar result regarding the larger values of $\lRe$ shown by AGNs is reported in \citet{Moral-Castro+2020} for isolated Sps. These authors attribute their outcome either to an internal angular momentum redistribution between the gas and the baryons remaining in the disc promoted by bars or other large-scale disc instabilities (e.g.\ spiral arms), or to external gas directly supplied to the central region of the galaxies through gas-rich minor mergers. Nevertheless, neither of these two scenarios appears well-suited for our Seyfert S0s, as they lack spiral arms, exhibit on average a low probability of bars in the MDLM-VAC-DR17 catalogue, and tend to inhabit rich environments (as indicated in the upper-right panel of Fig.~\ref{fig:cdf_mstar_density}). Therefore, to embrace the idea that the evolutionary path of certain Seyfert S0 may be related to RPS, it is necessary to consider other alternatives, such as collisions between the IGM and any residual ISM that may remain in the disks of these galaxies. This could potentially reduce the angular momentum of the latter, causing the inflow of a part of it towards the centre and consequently triggering nuclear activity.

\subsection{Activity driven by gravity}
\label{SS:gravity}
Lately, minor merger events are frequently invoked as drivers of both extended star formation activity in S0 galaxies located in the field or in small groups \citep[e.g.][]{mckelvie+2018,Deeley+2020,Rathore+2022,Coccato+2022,Maschmann+2022}, and localized star formation in inner \Ha\ rings \citep{Tous+2023}. The prevalence of these SF rings in diverse environments, including dense regions, suggests that they may originate from mergers with tiny neighbour dwarfs, as these are the only satellites capable of remaining trapped in the potential well of primaries during their infall into more massive haloes. Given that differences in the mass ratios of the merging galaxies may be associated with differences in the structure of activity induced in the remnants, we discuss below regular minor mergers and less conventional captures of small dwarf satellites, or mini mergers, separately.

\subsubsection{Minor mergers}
\label{SSS:minor_mergers}
In a recent study, \citet{Rathore+2022} concluded that the most plausible explanation for their observation, where more than half of the galaxies in a sample of 120 low-mass S0s with global SSFRs akin to SF Sps display signs of galaxy interactions, is rejuvenation through minor mergers. The central dominance of star formation in their SF lenticular galaxies aligns them well with our AC:OI--S0s, which not only feature activity profiles with the appropriate shape, but also exhibit star formation levels comparable to those observed in MS LTGs (see Fig.~\ref{fig:dps_dms_quenching}). Considering that our \DPS\ parameter serves as a reliable proxy for the SSFR in SF systems (Fig.~\ref{fig:bpt_dps_dms}), one can observe that the parallelism with the SF S0s studied by \citeauthor{Rathore+2022} further encompasses the dependence of the SSFR radial profiles on $M_\ast$ (compare the first row of panels in our Fig.~\ref{fig:dps_profiles_prop} with their fig.~7). Negative gradients are evident in both studies that become more pronounced as mass increases. These authors also find that although the SF S0s and Sps show similar SSFRs in their central regions, the radial gradients of the latter are positive on average. This reveals important differences with respect to the structural component, bulb or disc, respectively, which concentrates most of the star formation \citep[see also][]{GonzalezDelgado+2016}. Additionally, the younger typical ages of the stellar populations of our AC:OI--S0s (Fig.~\ref{fig:cdf_age}), along with their greater structural roundness and slightly more pressure-supported discs (Fig.~\ref{fig:lambda_epsilon}), are characteristics that fit perfectly into a rejuvenation scenario driven by minor mergers. 

As Fig.~\ref{fig:box_percentile_mstar} illustrates, the stellar mass distributions of S0s with OI activity gradients, primarily associated with the AC spectral class, exhibit a bias toward lower masses in the first two terciles of local density. Notably, in the second density tercile, where minor mergers are expected to be frequent, there is no overlap between the interquartile range of the OI subset (depicted by light brown boxes in the left panel) and those of the other two gradient types. If these galaxies originate from rejuvenated PS lenticulars with F or IO profiles, the relative paucity of high-mass OI systems at low and intermediate environmental densities might suggest that minor mergers are particularly effective in rejuvenating  low-mass passive S0s. This could be attributed to the higher ratio of residual gas to stars expected in these objects compared to their more massive counterparts. For the latter, the most likely outcome of such interactions would be a SF ring localized in the inner region of the disc, giving rise to an IO activity profile (see Section~\ref{SSS:mini_mergers}). As discussed in the preceding section, some high-mass AC:OI--S0s might originate from alternative mechanisms, such as RPS.

S0s can also form in one of the last steps of a chain of sequential minor mergers that feed the main progenitor and gradually change its morphology from a late spiral to an ETG system, as in the stream-driven merger scenario for galaxy formation discussed in \citet{Dekel+2009}. This form of evolution differs from the more dramatic impact of major mergers, which can produce more radical alterations of morphology, and it aligns well with the concept of hierarchical bulge growth described by \citet{Bournaud+2007}. It is also the preferred scenario by \citet{Maschmann+2022} to explain the central star-formation enhancements observed in a large sample of double-peak (DP) emission-line bulge-dominated LTGs, Sas and S0s identified from the SDSS. As noted by these authors, the possible merger-induced central star formation in DP galaxies occurs without a simultaneous increase in AGN activity.

One of the ways in which minor mergers involving gas-rich satellites can lead to the transition of Sps into S0s is by inducing galaxy-wide suppression of star formation. As revealed in the simulations conducted by \citet{Osman+2017}, the existence of counter-rotating gas in stellar discs has the potential to weaken or entirely inhibit gas density-enhancing structures like spiral arms and bars. In these circumstances, star formation is severely suppressed and can only proceed at a rate considerably lower than in co-rotating discs with comparable gas density. Notably, we find that the median SFR of the AC:OI--S0s located in the 'counter-rotation zone' of the anisotropy diagram (i.e., below the $\beta = 0.7 \varepsilon_{\rm intr}$ guideline) in the bottom-left panel of Fig.~\ref{fig:lambda_epsilon} is slightly lower than that of galaxies situated above it: $\log(\mbox{SFR}/(M_\odot\,\mbox{yr}^{-1})) = -0.58$ vs $-0.44$, respectively. Although this might be an overly simplistic approach to evaluate the feasibility of this scenario, the scant inferred difference still corresponds with the results of the simulations performed by \citeauthor{Osman+2017}, suggesting that a small fraction of our S0 galaxies could have followed this formation route. On the other hand, dry minor mergers may trigger morphological quenching \citep{Martig+2009}, a concept that involves the cessation of star formation in discs that become stable against fragmentation into bound gas clumps in galaxies that transition to a spheroid-dominated state after merging.

Late, gas-rich mergers with lower mass ratios (i.e., intermediate and/or major mergers) are more likely to give instead completely ellipsoidal SF systems, akin to the 15 blue SF elliptical galaxies identified by \citet{Lacerna+2020} in a sample of 343 objects of this morphology also extracted from the MaNGA survey. These SF ETGs exhibit younger stellar ages, lower metallicities, and essentially flat median SSFR profiles up to 1.5 \Reff. Although the limited number of such objects complicates the identification of general evolutionary trends, the authors of the study propose that these galaxies could be the result of recent wet mergers that rejuvenate intermediate- and low-mass classical Es in low or average density environments \citep[see also][]{Thomas+2010}. This idea represents a natural extension of the formation scenario proposed for our AC:OI--S0s, in which the mergers would be energetic enough to disrupt any preexisting disc and the initial mass of the satellite large enough to allow a substantial fraction of its core to reach the innermost regions of the primary \citep{Walker+1996}. The flatness observed in the SSFR radial gradients of the SF Es could then be a consequence of an increased central star formation that spans a significant portion of the remnants.

\subsubsection{Mini mergers}
\label{SSS:mini_mergers}
Numerical simulations aimed at investigating the formation histories of lenticular galaxies, such as those conducted by \citet{Deeley+2021}, reveal that most S0s formed through mergers undergo a phase wherein they develop a transient SF ring, typically lasting $\sim 1$--$2\,$Gyr. These rings consistently emerge in connection with minor merger events, their onset being preceded by the exhaustion of the gas in the central region of the hosts. They are often accompanied by the formation of a stellar bar or another non-axisymmetric gravitational distortion of the disc, signifying their close association with the central galaxy and, consequently, their coplanarity. Our examination of this phenomenon using MaNGA data, as detailed in \citet{Tous+2023}, has revealed that the existence of inner ($\langle R\rangle\sim 1\,$\Reff) SF rings in S0 galaxies is largely associated with an IO configuration of their activity gradients. Additionally, our study has confirmed the ubiquity of these rings, particularly among massive S0s, with fractional abundances reaching $\sim 50$ per cent independently of both their spectral type (AC or PS) and environmental density. Drawing from these findings, we deduced that the minor mergers responsible for the formation of these SF rings probably involve tiny satellites closely bound to the primary galaxy.

Regarding the possibility that these small satellite captures could also drive AGN activity, simulations such as those conducted by \citet{Mapelli+2015}, which explore the build up of gas rings in S0s through minor mergers of mass ratio $\sim 1$:$20$, show that while much of the stars and gas initially bound to the satellites are scattered into the halo of the central galaxy, a small fraction of their material can be channelled towards its nucleus. However, it is important to note that in these simulations satellites were also modelled as disc galaxies equipped with a stellar bulge. In a mini-merger scenario, where most merging satellites are expected to have a more diffuse stellar structure of the Irr or Sph type, they are less likely to reach the host's nucleus before being fully disrupted, unless their initial orbit is well aimed towards this region \citep{Kendall+2003}. 

Our analysis of the IO radial activity profiles aligns seamlessly with the scenario outlined above. The second and fifth rows of panels in Fig.~\ref{fig:dps_profiles_prop} show that the median IO profiles exhibit a more pronounced S-shape with increasing stellar mass, a behaviour that becomes less prominent when these profiles are represented as a function of environmental density. There is also the substantial overlap shown by the interquartile ranges of the mass distributions of the IO and F subsets across all density intervals (depicted, respectively, by the orange and brown boxes in Fig.\ref{fig:box_percentile_mstar}), which supports the notion that the former could be rejuvenated versions of the latter. The rejuvenation scenario is further supported by the observation that, within each spectral class, the stellar populations of S0s with IO profiles exhibit statistically younger luminosity-weighted ages than  those of the other gradients (see the left panels of Fig. \ref{fig:cdf_age}). However, the fact that, as shown in Fig.\ref{fig:dps_dms_quenching}, lenticulars with this type of activity profile typically reside in the green valley (GV), suggests that these rejuvenation events trigger localized, bursty star formation episodes that significantly increase the SFRs of the galaxies, but have a less pronounced effect on their SSFRs compared to S0s with OI profiles.

\subsection{Is the PS:F subset the S0s' graveyard?}
\label{SS:graveyard}
The relatively short-lived and transient nature of both the rejuvenation and fading phenomena implies that all the evolutionary pathways of S0s are expected to eventually lead to quiescent systems characterized by a fundamentally bland activity profile. In our classification system, these features correspond to S0s of spectral class PS and F-type gradients. Consequently, it is natural to expect that this subset of galaxies has properties compatible with the notion of attractors of evolutionary histories.

Certainly, the PS:F subset stands out as the largest within the MaNGA sample (Table~\ref{tab:sample_sizes}). This, coupled with the observation that these galaxies also host the oldest stellar populations (Fig.~\ref{fig:cdf_age}), indicates that they are the most prevalent and enduring state among present-day S0s. Furthermore, the distribution of their stellar masses spans the entire range of values seen in the other subsets, while the frequency distribution of their environmental densities encompasses all percentile ranks (Figs.~\ref{fig:cdf_mstar_density} and \ref{fig:box_percentile_mstar}). To all this must be added the fact that the PS:F systems are the only ones that show a distinctly bimodal distribution in the $(\lRe,\varepsilon)$ diagram (Fig.~\ref{fig:lambda_epsilon}), with two peaks that differ basically in the value of the ellipticity. Assuming that the positioning of galaxies in this diagram reflects the evolutionary path they have followed, the mode centred on the lowest value of $\varepsilon$ would be associated with galaxies that, after attaining the S0 status, have undergone one or more transient episodes of rejuvenation during which their SFRs have been considerably intensified (as evident from the comparable positions of the left-hand peak and the peak of the SF AC subset). Our study's findings suggest that minor mergers could be the most effective catalysts for these periods of heightened activity. Such mergers would result in rounder objects by triggering disc instabilities and heating, although, as deduced from the anisotropy diagrams, they would barely impact the rotational support of the primary galaxy. On the other hand, the right-hand mode would correspond to galaxies that have experienced a less eventful formative history, in which once they have become fully-formed S0s have not subsequently suffered significant structural and dynamic alterations.

\section{Summary and conclusion}
\label{S:summary}
We have studied the radial activity profiles up to $1.5\,$\Reff\ of 1072 present-day S0 galaxies drawn from the SDSS-IV MaNGA survey as a function of their mass, age, structure and kinematics, and density of their environment to see how they fit with the different evolutionary histories proposed for these objects. The activity profiles have been obtained after a process of dimensionality reduction consisting in projecting the composite spectra that result from stacking the spaxels of the data cubes, grouped in radial bins, into the PC1--PC2 space defined by their first two principal components, which encompass most of the variance of the whole sample of spectra. In this latent space, three zones or spectral classes can be distinguished that correlate with the level of activity of the galaxies and that we identify as the passive sequence (PS class), the active cloud (AC), and the intermediate transition region (TR). Near 90 per cent of the PC1--PC2 representations of the radial profiles are well approximated by straight lines, which has facilitated the division of the above spectral classes into outside-in (OI), inside-out (IO) and flat (F) categories according to, respectively, the negative, positive and null value of the radial activity gradient. The introduced classifications based on the \DPS\ parameter, as well as on its radial gradient, provide a unified picture of the spectral information of S0 galaxies in the optical range with respect to their overall activity level and, in combination with more physical quantities and line diagnostics, help to interpret key trends in these systems. Next, we provide a summary of the main findings of this study.

\begin{itemize}
    \item The gradients of the radial activity profiles of S0s show a close relationship with their spectral class, BPT type and SSFR.
    
    \item PS--S0s very often exhibit low-level, F-type activity gradients as well as a significant number of IO profiles, while in a substantial fraction of AC--S0s activity increases OI. F activity gradients in AC lenticulars and OI in PS systems are uncommon.

    \item Most SF and Seyfert S0s are AC systems that preferentially show OI activity gradients, while a good number of LINER S0s fall in our PS spectral class. The fact that the vast majority of the latter show F or IO activity gradients suggests that they are weakly-active systems likely powered by post-AGB stars. Among the S0s with a PS status, we also identify 9 objects with E+A-type spectra located in the MS of star formation. 
    
    \item A significant portion of S0s in the MS exhibit negative activity gradients, whereas among the quiescent ones, the majority have flat gradients. Positive activity gradients are frequently observed in GV objects.

    \item The significance of recent star formation in relation to overall activity is greater in S0 galaxies displaying IO profiles compared to their OI and F counterparts, irrespective of the activity level.
    
    \item Among low-mass AC--S0s, the highest levels of activity correspond to objects classified as SF. However, the progressive reduction of the SSFR with stellar mass makes Seyfert S0s to become the most active systems at intermediate and high masses. We also observe that star formation tends to be less intense when AGN activity dominates, which suggests negative feedback. 
    
    \item The greatest abundance of SF AC--S0s compared to those classified as Seyfert leads to the decrease of the activity level in this spectral class with stellar mass at all galactocentric radii, an effect that is particularly noticeable in galaxies with OI activity profiles. This behaviour, which is consistent with the SSFR--$M_\ast$ relationship defining the MS, is however inverted in the PS--S0s with IO profiles, pointing to a faster increase of the star formation efficiency with stellar mass in these latter objects than in typical SF galaxies.

    \item The ratio between positive and negative activity gradients in AC--S0s increases with stellar mass. 

    \item The relationship between the median level of the activity profiles of S0 galaxies and the local density of their environment is, in general, less clear-cut than with stellar mass, showing only indications in the highest density bin of a slight suppression of the intensity in the AC:OI systems and of a similarly light increase in it in the few objects of this same spectral class with IO gradients that inhabit these richer environments.
 
    \item The bivariate distributions of S0s in the $(\lRe,\varepsilon)$ anisotropy diagram show that AC:OI systems classified as SF are rounder and slightly more pressure supported, and their Seyfert counterparts more rotationally supported, than the bulk of the local population of these objects. For its part, the distribution of PS--S0s is bimodal, with a peak centred around values typical of lenticular systems, along with another one that roughly aligns with that of the SF subset.

\end{itemize}

The approach followed in this work is  complementary to those followed in similar efforts that focus on the spatial gradient of properties exclusively related to star formation. While the results of our analysis are compatible with the different evolutionary scenarios proposed in the literature for fully-formed S0s, they also reveal a certain tendency to identify gravitational encounters, more specifically, two different types of dry minor mergers, as the primary actors, even in rich environments, at the expense of more traditional mechanisms involving hydrodynamic interactions. 

In forthcoming papers, we will extend this methodology to the entire Hubble sequence while incorporating information about the third principal component. The latter, in addition to increasing the fraction of explained sample variance, will allow us to self-consistently disentangle star formation from both nuclear accretion and photoionization by post-AGB stars in any galaxy from a sample of spatially resolved spectra, thus avoiding the use of external activity diagnostics to identify the source of ionization. 

\section*{Acknowledgements}
We acknowledge financial support from the Spanish state agency MCIN/AEI/10.13039/501100011033 and by 'ERDF A way of making Europe' funds through research grants PID2019--106027GB--C41 and PID2019--106027GB--C43. MCIN/AEI/10.13039/501100011033 has also provided additional support through the Centre of Excellence Severo Ochoa's award for the Instituto de Astrof\'\i sica de Andaluc\'\i a under contract SEV--2017--0709 and the Centre of Excellence Mar\'\i a de Maeztu's award for the Institut de Ci\`encies del Cosmos at the Universitat de Barcelona under contract CEX2019--000918--M. J.L.T.\ acknowledges support by the PRE2020--091838 grant from MCIN/AEI/10.13039/501100011033 and by 'ESF Investing in your future'. H.D.S.\ acknowledges support by the PID2020-115098RJ-I00 grant from MCIN/AEI/10.13039/501100011033  and from the Spanish Ministry of Science and Innovation and the European Union - NextGenerationEU through the Recovery and Resilience Facility project ICTS-MRR-2021-03-CEFCA. This project makes use of the MaNGA-Pipe3D dataproducts. We thank the IA-UNAM MaNGA team for creating this catalogue, and the Conacyt Project CB-285080 for supporting them.

\section*{Data availability}
This research has made prominent use of the following
databases in the public domain: the SDSS Science Archive Server (\url{https://data.sdss.org/sas/}), the VizieR Online Data Catalog J/MNRAS/515/3956 (\url{https://cdsarc.cds.unistra.fr/viz-bin/cat/J/MNRAS/515/3956}) and the GSWLC-2 catalogue (\url{https://salims.pages.iu.edu/gswlc/}).
%%%%%%%%%%%%%%%%%%%%%%%%%%%%%%%%%%%%%%%%%%%%%%%%%%

%%%%%%%%%%%%%%%%%%%% REFERENCES %%%%%%%%%%%%%%%%%%

\bibliographystyle{mnras}
\bibliography{biblio}

%%%%%%%%%%%%%%%%%%%%%%%%%%%%%%%%%%%%%%%%%%%%%%%%%%

%%%%%%%%%%%%%%%%% APPENDICES %%%%%%%%%%%%%%%%%%%%%
\appendix
\section{Robustness of the results against morphological contamination}
\label{A:morphological_contamination}
In this appendix, we test the robustness of our results against biases caused by uncertainties in the selection of the lenticular morphology. As explained in Section~\ref{S:data}, to identify the S0s studied in this work we have adopted the selection criteria $T < 0.5$, as well as the probabilities $P_{\mathrm{LTG}} < 0.5$ and $P_{\mathrm{S0}}\geq 0.5$ recommended in the Deep Learning-based morphological catalogue of \citet{HDS+2022}. Additionally, we have thoroughly reviewed the SDSS images to eliminate spiral contaminants that might have bypassed these filters. However, it is important to note that when probabilities are close to the limits mentioned above, the likelihood that a galaxy is an S0 becomes comparable to the likelihood that it is actually an elongated E or an Sp with faint spiral arms. To evaluate the effects of this potential bias in our results, we now increase the priority given to purity over completeness in our data by placing yet more stringent constraints on the sample selection. Thus, we have repeated our analysis on a subsample of MaNGA S0 galaxies obtained by replacing the conditions stated above by the more restrictive selection criteria $P_{\mathrm{LTG}} < 0.3$ and $P_{\mathrm{S0}} \geq 0.7$. We consider this approach to be preferable to other options, such as adopting a minimum value in the specific angular momentum and ellipticity of the galaxies to minimise the presence of E systems, since measurements of these two magnitudes are affected by large uncertainties, or imposing an upper limit on the inclination or on the SSFR that minimises contamination by SF Sps, but that could also remove S0s with significant star formation.

% Fig. A1
\begin{figure}
    \centering
	\includegraphics[width=\columnwidth]{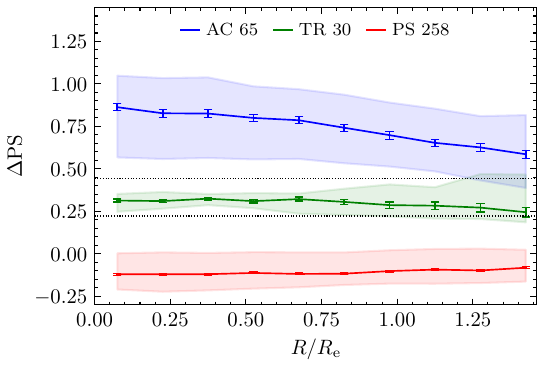}
 \vspace{-0.6cm}
    \caption{As Fig.~\ref{fig:dps_profiles} but for the reduced sample of S0s defined with a stricter morphological selection criteria (see text for details) and without including the spectral profiles of AC galaxies classified as SF and Seyfert.}
    \label{fig:dps_profiles_pure}
\end{figure}

% Fig. A2
\begin{figure}
    \centering
	\includegraphics[width=0.95\columnwidth]{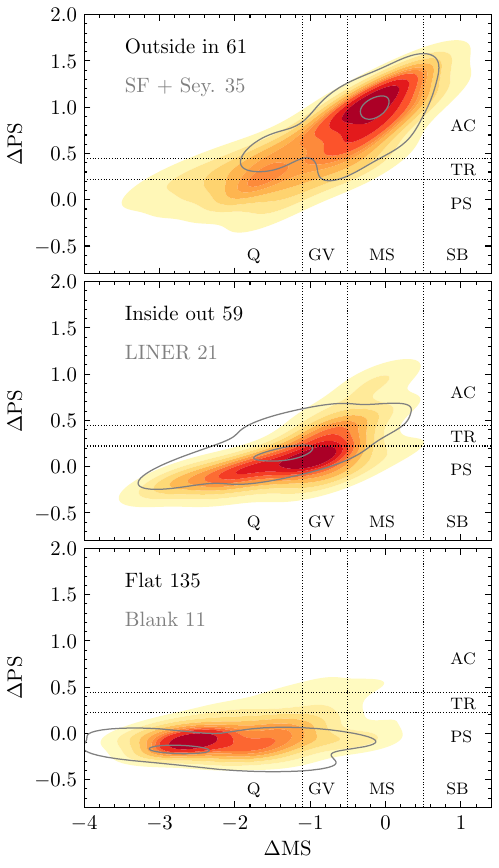}
\vspace{-0.3cm}
        \caption{As Fig.~\ref{fig:dps_dms_quenching} but for the reduced sample of S0s defined with a stricter morphological selection criteria (see text for details).}
\label{fig:dps_dms_quenching_pure}
\end{figure}

After applying the above stricter selection criteria to the MaNGA catalogue, we are left with a sample of $353$ S0s. The significantly smaller size of this dataset (contains about one third of the original S0s) makes it impossible to repeat each and every one of the analyses carried out in the main text of this work, but it does allow us replicating an important part of them, enough to confirm that the results obtained with the original sample are not affected by any substantial morphological bias.

\begin{figure}
    \centering
    \includegraphics[width=0.84\columnwidth]{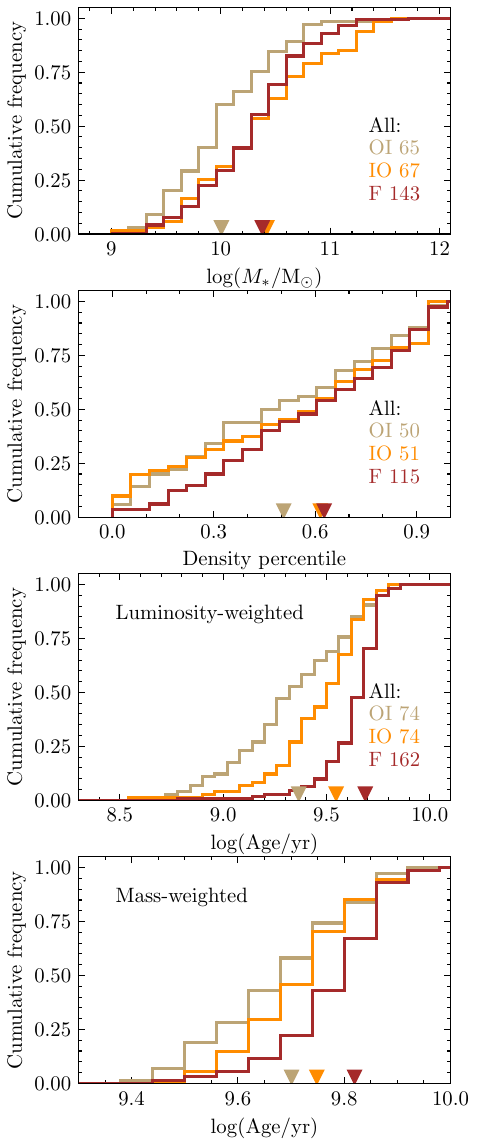}
    \vspace{-0.3cm}
    \caption{\emph{From top to bottom:} cumulative distribution functions of the stellar mass, local density of the environment expressed by percentiles, and luminosity- and mass-weighted ages for the reduced sample of S0s defined with a stricter morphological selection criteria (see text for details). The histograms represent the CDFs of galaxies with outside in (OI, light brown), inside out (IO, orange), and flat (F, brown) activity gradients. The inverted filled triangles show the medians of these CDFs. The size of each subset is indicated in the legend next to the profile labels. To be compared with the bottom panels of Figs.~\ref{fig:cdf_mstar_density} and \ref{fig:cdf_age}.}
    \label{fig:cdfs_pure}
\end{figure}

\begin{figure*}
    \centering
    \includegraphics[width=0.94\textwidth]{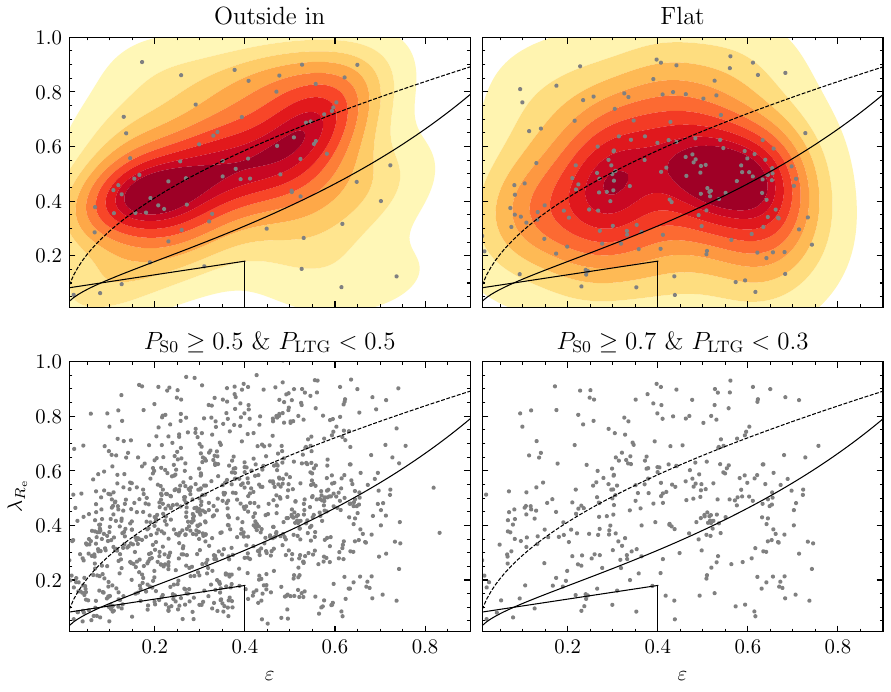}
    \caption{Upper panels are as the upper panels of Fig.~\ref{fig:lambda_epsilon} but for the reduced sample of S0s defined with a stricter morphological selection criteria (see text for details). Lower panels show the individual $(\lRe,\varepsilon)$ values for the 1072 S0s studied in this work (left) and for the 373 S0s in the reduced sample (right). Comparison of the two scatter diagrams demonstrates that the data loss experienced by the reduced sample is considerably greater around the slow rotator area than in the rest of the diagram.}
    \label{fig:lambda_epsilon_pure}
\end{figure*}

Thus, as shown by Fig.~\ref{fig:dps_profiles_pure}, the median radial \DPS\ profiles of the different spectral classes inferred from the reduced sample of S0s turn out to be nearly identical to those derived in Section~\ref{S:activity_profiles} for the initial sample and depicted in Fig.~\ref{fig:dps_profiles}. Likewise, the \DMS--\DPS\ diagrams in Fig.~\ref{fig:dps_dms_quenching_pure} reveal that the tight relationships between the distributions of activity gradients, spectral and BPT classes, and star formation status of the galaxies featured in Fig.~\ref{fig:dps_dms_quenching} also continue to hold perfectly. The aggregated CDFs of stellar mass, environmental density, and luminosity- and mass-weighted stellar ages of the different activity gradients depicted in the four panels of Fig.~\ref{fig:cdfs_pure} maintain clear consistency with their corresponding previous results as well (see the bottom panels of Fig.~\ref{fig:cdf_mstar_density} and \ref{fig:cdf_age}), thus preserving the hierarchy of the mean values of these variables established by the original data in Section~\ref{S:profiles_vs_properties}. In the same way, the general shapes of the aggregated bivariate distributions of the different spectral classes and gradient types in the $(\lRe,\varepsilon)$ diagram are also preserved. This allows us to identify in the two upper panels of Fig.\ref{fig:lambda_epsilon_pure} the same trends between the observed ellipticity and specific angular momentum of stars shown by the more complete samples of S0s with OI and F activity gradients represented in Fig.\ref{fig:lambda_epsilon}. Additionally, the significant reduction in the number of galaxies due to the more stringent morphological selection reveals an interesting effect related to the kinematics of these objects. As the lower panels of Fig.~\ref{fig:lambda_epsilon_pure} show, this 'morphological cleaning' has a particularly noticeable impact in the region of slow rotators, which reduce their presence in the new sample to only $2.5$ per cent, when before this fraction was around $6$ per cent. This supports \citet{Capellari2016}'s claim that slowly rotating E galaxies can be differentiated from S0 galaxies using just their images.

% Fig. A3

% Fig. A4
%%%%%%%%%%%%%%%%%%%%%%%%%%%%%%%%%%%%%%%%%%%%%%%%%%

% Don't change these lines
\bsp	% typesetting comment
\label{lastpage}
\end{document}